\documentclass[%
 aip,
 amsmath,amssymb,
 reprint,%
]{revtex4-1}

\usepackage{graphicx}% Include figure files
\usepackage{dcolumn}% Align table columns on decimal point
\usepackage{bm}% bold math
\usepackage[utf8]{inputenc}
\usepackage[T1]{fontenc}
\usepackage{mathptmx}
\usepackage{etoolbox}
\usepackage{subfig}
\usepackage[hidelinks]{hyperref}
\usepackage[capitalise,nameinlink]{cleveref}
\crefname{figure}{Fig.}{Figs.}
\Crefname{figure}{Fig.}{Figs.}

\usepackage{graphicx}
\usepackage{subcaption}
\usepackage{color}

\usepackage{placeins}  % \FloatBarrier
\makeatletter
\def\@email#1#2{%
 \endgroup
 \patchcmd{\titleblock@produce}
  {\frontmatter@RRAPformat}
  {\frontmatter@RRAPformat{\produce@RRAP{*#1\href{mailto:#2}{#2}}}\frontmatter@RRAPformat}
  {}{}
}%
\makeatother
\begin{document}

\preprint{AIP/123-QED}

\title{Energy Landscape Structure of Small Graph Isomorphism Under Variational Optimization}
% Force line breaks with \\
\author{Turbasu Chatterjee}
\affiliation{ 
 Department of Computer Science and Engineering, Calcutta Institute of Engineering and Management,\\
 Maulana Abul Kalam Azad University of Technology, Kolkata, India%\\This line break forced with \textbackslash\textbackslash
}%
\author{Shah Ishmam Mohtashim}%
 \email{sishmam51@gmail.com}
\affiliation{ 
Department of Chemistry, University of Dhaka, Dhaka, Bangladesh%\\This line break forced with \textbackslash\textbackslash
}%

\author{Akash Kundu}
 % \homepage{http://www.Second.institution.edu/~Charlie.Author.}
\affiliation{%
 Institute of Theoretical and Applied Informatics,
	Polish Academy of Sciences, Bałtycka 5, 44-100 Gliwice, Poland.} 
	
\altaffiliation[Also at ]{Joint Doctoral School, Silesian University of Technology, Akademicka 2a, 44-100 Gliwice, Poland}

\date{\today}% It is always \today, today,

\begin{abstract}
We investigate a quadratic unconstrained binary optimization (QUBO) formulation of the graph isomorphism problem using the Quantum Approximate Optimization Algorithm (QAOA) and the Variational Quantum Eigensolver (VQE). For small graph instances, we observe that isomorphic pairs exhibit consistent clustering in variational energies, indicating that the Hamiltonian successfully encodes structural features. However, we demonstrate that low variational energy alone is an unreliable certifier of isomorphism due to the high probability of converging to infeasible states that violate bijection constraints. To address this, we analyze optimization trajectories rather than final energies; consistently outperform naive energy thresholding, though absolute performance remains limited. Our results characterize the current limits of variational algorithms for graph isomorphism, positioning energy landscape analysis as a diagnostic tool rather than a scalable decision procedure in the NISQ regime.

\end{abstract}

\maketitle

% \begin{quotation}
% The ``lead paragraph'' is encapsulated with the \LaTeX\ 
% \verb+quotation+ environment and is formatted as a single paragraph before the first section heading. 
% (The \verb+quotation+ environment reverts to its usual meaning after the first sectioning command.) 
% Note that numbered references are allowed in the lead paragraph.
% %
% The lead paragraph will only be found in an article being prepared for the journal \textit{Chaos}.
% \end{quotation}

\section{\label{sec:level1}Introduction}

The graph isomorphism problem holds a special place in theoretical computer science because it sits in a gray area of computational complexity. While not known to be NP-complete, it is also not proven to be solvable in polynomial time, and Babai’s breakthrough result \cite{babai2016graph} established a quasi-polynomial time classical algorithm — a landmark in complexity theory. Beyond its foundational role, the problem appears in diverse areas such as molecular similarity in chemistry, protein interaction networks in biology, and structural equivalence in social networks \cite{shehab24, rinq25}. These applications underscore both the theoretical depth and the practical importance of efficient quantum approaches.

Hybrid quantum-classical algorithms have emerged as a leading strategy to obtain quantum advantage in Noisy Intermediate Scale Quantum (NISQ) devices \cite{kyriienko25, cerezo25, glass25, 10313751, trev_ieee, 2021, Preskill2018quantumcomputingin}. In particular, variational quantum algorithms (VQAs), which provide a general framework for tackling important classes of problems, involve quantum circuits whose parameters are are optimized by classical optimisation techniques \cite{pulse21}. Typically in VQAs, like the Variational Quantum Eigensolver (VQE) \cite{peruzzo2014} and Quantum Approximate Optimization Algorithm (QAOA) \cite{farhi2014quantum}, the goal is to find a quantum state which minimizes the energy of a problem Hamiltonian. Earlier, quantum annealing (QA) had enabled quantum-enhanced information processing for roughly the same areas of application as VQAs \cite{10.3389/fphy.2014.00005}. The key difference between VQAs and QA is that the former involves a gate-based computation, and the latter is quantum-fluctuation-based computation exploiting quantum mechanical effects such as tunneling and entanglement. Quantum Annealers are relatively easier to build and have significantly more qubits than gate-based quantum devices \cite{2018}.

What underpins quantum annealing is the adiabatic theorem in physics, which involves slowly evolving the ground state of a physical system from the problem Hamiltonian of the system, and is what enables finding the ground state energies relevant to QUBO problem formulations \cite{Santoro_2006, lewis2017quadratic, domino2021quadratic}. Most prior quantum information efforts to tackle the graph isomorphism problem have focused on quantum annealing approaches. While annealers are effective for specific instances, they face limitations in scalability, connectivity, and flexibility. In contrast, VQAs provide greater adaptability to different problem formulations on account of relying on a gate-based circuit model. They are furthermore empowered by the hybrid quantum-classical optimization loop, making them more robust for implementations on noisy intermediate-scale quantum (NISQ) devices. This work presents an exploratory small-instance, proof-of-principle study within a NISQ framework, complementing earlier quantum annealing approaches and highlighting optimization behaviour and limitations. While quantum annealing has proven effective for small instances of graph isomorphism, it faces several intrinsic limitations. First, annealers rely on fixed hardware connectivity graphs, requiring costly minor embedding for dense or irregular QUBO instances, which can significantly inflate qubit overhead and degrade solution quality. Second, the annealing schedule and mixer dynamics are largely fixed, limiting flexibility in exploiting problem-specific structure or symmetries. Third, annealers primarily provide access to the ground state, offering limited insight into the broader energy landscape that may encode useful structural information.

Gate-based variational algorithms address these limitations by enabling fully programmable Hamiltonians, mixers, and ansätze, allowing greater flexibility in encoding constraints and symmetries. Moreover, variational methods provide access not only to ground-state energies but also to the structure of low-energy excitations and energy clustering phenomena. This flexibility is particularly valuable in the NISQ era, where error mitigation, adaptive circuit design, and hybrid quantum-classical workflows can be tailored to specific problem instances. As circuit-model quantum hardware continues to scale, variational approaches offer a natural pathway toward more expressive and scalable quantum heuristics for graph isomorphism.

By considering VQAs for tackling graph isomorphism, this work explores whether the variational paradigm offers not only practical performance on small graph instances, but also a potential pathway to scalable quantum heuristics for one of the most enigmatic problems in complexity theory.

\subsection{Related work}

Quantum approaches to the Graph Isomorphism (GI) problem have historically followed two largely distinct directions. The first line of work attempts to cast GI as an instance of the non-Abelian Hidden Subgroup Problem (HSP), solvable in principle via quantum Fourier sampling. Early optimism surrounding this approach was tempered by a series of negative results demonstrating that standard coset-state and Fourier-sampling techniques fail to efficiently distinguish automorphism groups, even for restricted families of graphs~\cite{10.1145/1857914.1857918,10.1145/380752.380769,shehab2017quantum}. As a consequence, the HSP-based approach is now widely regarded as unlikely to yield a general and efficient quantum algorithm for graph isomorphism.

A second, more practically successful direction encodes GI as a quadratic unconstrained binary optimization (QUBO) or Ising Hamiltonian, solvable via quantum annealing. Seminal works by Lucas~\cite{10.3389/fphy.2014.00005}, Zick \textit{et al.}~\cite{2015}, and Calude \textit{et al.}~\cite{calude2017qubo} demonstrated that the GI problem can be mapped to an energy minimization task in which valid isomorphisms correspond to ground states of a suitably constructed Hamiltonian. These formulations have become the dominant quantum heuristic for GI and align with the broader success of QUBO-based models for combinatorial optimization problems such as Boolean satisfiability and the Traveling Salesman Problem~\cite{farhi2000quantum,salehi2021unconstrained}.

Despite their effectiveness, annealing-based approaches face intrinsic hardware-level limitations. The dense logical connectivity required by GI QUBOs must be embedded into sparse physical hardware graphs, often necessitating complex minor-embedding procedures. This embedding overhead substantially inflates the number of required physical qubits relative to the logical problem size, thereby limiting scalability on current annealing platforms. Furthermore, standard annealers employ fixed driver Hamiltonians, restricting the flexibility with which the energy landscape can be explored.

In contrast, comparatively little work has investigated graph isomorphism using \emph{gate-based variational quantum algorithms} (VQAs), such as the Quantum Approximate Optimization Algorithm (QAOA) or the Variational Quantum Eigensolver (VQE). To the best of our knowledge, there are no prior demonstrations applying such circuit-model variational methods directly to the GI problem. This absence is largely explained by resource considerations: a natural QUBO encoding of GI for graphs of order $n$ requires $O(n^2)$ logical qubits, a scale that has only recently become accessible for small instances on gate-based quantum hardware and high-fidelity simulators. Historically, annealing architectures provided the only platforms with sufficient qubit counts to explore such encodings, even at modest problem sizes.

The motivation for revisiting GI within a gate-based variational framework is therefore complementary rather than competitive with quantum annealing. While annealers excel at large-scale ground-state search within fixed Hamiltonian families, variational circuit-model approaches offer distinct advantages in terms of flexibility and analytical access. Logical connectivity can be handled via compilation techniques, such as SWAP networks, rather than physical embedding, potentially mitigating qubit overhead at the expense of circuit depth. Moreover, variational algorithms allow for programmable ansätze, non-commuting mixer Hamiltonians, and access to intermediate optimization dynamics, features that are largely inaccessible in standard annealing paradigms. We emphasize that this work does not claim a computational advantage over classical algorithms for graph isomorphism, but instead focuses on characterizing the behavior and limitations of variational quantum algorithms applied to isomorphism-encoding Hamiltonians in the small-graph regime.

Recent theoretical results further motivate this perspective. Szegedy~\cite{szegedy2019qaoa} showed that QAOA expectation values at fixed depth define graph invariants, suggesting that variational energy landscapes may encode nontrivial structural information beyond strict ground-state optimization. This observation motivates an exploratory study of how graph structure manifests in variational energy landscapes, even when exact isomorphism certification is infeasible.

Accordingly, the present work should be viewed as a proof-of-principle investigation of graph isomorphism within the gate-based variational paradigm, restricted to small graphs in the NISQ-accessible regime. Rather than proposing a scalable solution to GI, we aim to characterize how isomorphism-encoding Hamiltonians behave under variational optimization, to identify both the capabilities and limitations of energy-based heuristics, and to clarify the role of variational dynamics in extracting structural information from quantum encodings of GI.

%%%%%%%%%%%%%%%%%%%%%%%%%%%%%%%%%
\
\section{\label{sec:level2}Background}

The Graph Isomorphism Problem is the computational problem of determining whether two finite graphs are structurally identical or isomorphic. An isomorphism of a graph $G = (V, E)$ to a graph $H = (W, F)$ is a one-to-one, bijective mapping from the vertex set of the first graph $V$ to the vertex set of the second graph $W$ that preserves adjacency and non-adjacency. The exact complexity of the problem has eluded researchers ever since its conception for more than five decades. The problem is not known to be solvable in polynomial time generally, but interestingly, for many special classes of graphs, the problem becomes amenable to polynomial time solutions \cite{10.1145/321864.321877}. If the graphs have different sizes or orders, then they cannot be isomorphic, and these cases can be decided quickly. The current best known classical algorithm for the problem has a time complexity of $\exp \left( (\log n)^{O(1)} \right)$ which is quasi-polynomial. \cite{babai2016graph} The quantum computational complexity of the problem is unknown \cite{shehab2017quantum}. Previous attempts using quantum algorithms in the light of Hidden Subgroup Problem, using Quantum Fourier Sampling has been shown to fail \cite{10.1145/1857914.1857918, 10.1145/380752.380769, 1998}.

In light of these negative results regarding the standard algebraic frameworks, recent attention has shifted towards heuristic optimization methods better suited for Near-Term Intermediate Scale Quantum (NISQ) devices. The following sections detail the theoretical foundations of the Quantum Approximate Optimization Algorithm (QAOA) and the Variational Quantum Eigensolver (VQE), which serve as the primary variational architectures for our proposed approach.

\subsection{The Quantum Approximate Optimization Algorithm}
QAOA, introduced by Farhi et.al \cite{farhi2014quantum}, finds an approximation to the ground state of a problem Hamiltonian by
constructing a specific variational ansatz through first order Suzuki-Trotter decomposition \cite{lancaster2025simulating, aplQ, Lee_2023} approximating adiabatic evolution. The operators $\textrm{exp}(- i \beta H_\text{m})$ and $\textrm{exp}(- i \alpha H_\text{o})$ are applied in
alternation resulting in the state
\begin{equation}
\lvert \mathbf{\alpha}, \mathbf{\beta}\rangle = \prod_{k = 1}^{L} \textrm{exp}(- i \beta_k H_\text{m}) \textrm{exp}(- i \alpha_k H_\text{o})\lvert + \rangle^{\otimes n},
\end{equation}
where $H_\text{o}$ is the objective and $H_\text{m} = -\sum_j X_j$ is the mixer Hamiltonian. Under fixed number of layers (i.e. $L$), the algorithm requires $2L$ parameters. The expectation
value $\langle \alpha, \beta | H | \alpha, \beta \rangle $, of state $\lvert \alpha , \beta \rangle$ is approximated through measuring the state in the
computational basis. The parameters $\alpha$, $\beta$ are updated using classical optimization procedures so that the energy returns an optimal value. QAOA can be harnessed to solve problems such as Max-Cut\cite{farhi2001quantum, majumdar2021optimizing}, Max k-Vertex Cover\cite{2020} and even integer factorization \cite{anschuetz2018variational}.

\subsection{The Variational Quantum Eigensolver}

The VQE uses Ritz's variational principle to prepare an approximation to the ground state of a Hamiltonian and to estimate its corresponding energy. In this algorithm, the quantum device is tasked with preparing parameterized trial states 
\[
\lvert\psi (\mathbf{\theta})\rangle = U(\mathbf{\theta}) \lvert 0\rangle,
\]
where $U(\mathbf{\theta})$ is a parameterized unitary depending on a set of classical parameters $\mathbf{\theta}$. The expectation value of the Hamiltonian, 
\[
E(\mathbf{\theta}) = \langle \psi (\mathbf{\theta}) | H | \psi (\mathbf{\theta}) \rangle,
\]
is measured on the quantum hardware, and a classical optimizer is employed to iteratively update $\mathbf{\theta}$ until the minimum is reached. The variational principle ensures that every evaluated energy provides an upper bound on the true ground state energy, making the method both flexible and robust. 

A key advantage of VQE over purely classical methods is its ability to efficiently explore quantum states that are exponentially hard to represent on a classical computer. This is particularly significant for problems where correlations or entanglement play a central role. Consequently, VQE has become one of the most widely studied near-term quantum algorithms, with applications spanning quantum chemistry, condensed matter physics, materials science, and combinatorial optimization \cite{plesch2024}. Despite its usefulness, VQE is computationally costly as each optimization step requires repeated circuit preparation and measurement, which quickly becomes expensive as system size grows. Moreover, the choice of ansatz and optimizer strongly affects performance, and noise in near-term devices adds additional challenges \cite{glass25, trev_ieee}. However, these costs are balanced by its unique strengths: (i) it provides a variational certificate of accuracy through Ritz's principle, (ii) it allows flexible tailoring of ansatz circuits to exploit problem structure, and (iii) it can act as a high-precision validator of phenomena first observed in lighter, more approximate algorithms such as QAOA. In the context of the Graph Isomorphism problem, the use of VQE is particularly valuable as a complementary check. Whereas QAOA reveals clustering in energy landscapes across isomorphic graph families at relatively low circuit depth, VQE confirms the same phenomenon with greater fidelity, albeit at a higher computational cost. This duality—fast approximate detection via QAOA and rigorous confirmation via VQE—makes VQE an essential component in our variational study, demonstrating that the observed clustering effect is not an artifact of shallow circuits but rather an intrinsic property of the underlying Hamiltonian encoding of isomorphism.

\section{\label{sec:level3}Formulation of quadratic program}

The first described baseline penalty Hamiltonian for the graph isomorphism was introduced in Lucas \cite{10.3389/fphy.2014.00005}. Let us consider $G_1 = (V_1, E_1)$, and $G_2 = (V_2, E_2)$ graphs, where $V$ are the vertices and $E$ edges. The Hamiltonian is formulated by proposing binary variables $x_{v,i}$, which is $1$ if the vertex $v$ in $G_2$ gets mapped to vertex $i$ in $G_1$ as follows:
\begin{widetext}
\begin{equation}
    H = A\sum_v \left( 1- \sum_i x_{v,i}\right)^2 + A\sum_i \left( 1- \sum_v x_{v,i}\right)^2 +B\sum_{ij \notin E_1 }\sum_{uv \in E_2 } x_{u,i}x_{v,j} + B\sum_{ij \in E_1 }\sum_{uv \notin E_2 } x_{u,i}x_{v,j}. \label{eq:lucas-hamil} 
\end{equation}
\end{widetext}
In Zick \textit{et.al} \cite{shehab2017quantum} a modified formulation of Eq.\ref{eq:lucas-hamil} is proposed in order to achieve a simple set of coupler values (e.g. {0, 1} instead of {0, 1,
2} in Lucas) amenable to quantum annealing. An edge related penalty is applied to a coupling iff there is not a vertex mapping penalty. The modified Hamiltonian is given as follows:

\begin{widetext}
\begin{equation}
\begin{split}
    H = A\sum_v \left( 1- \sum_i x_{v,i}\right)^2 + A\sum_i \left( 1- \sum_v x_{v,i}\right)^2 +B\sum_{\substack{ij \notin E_1 \\ i \neq j }}\sum_{uv \in E_2 } x_{u,i}x_{v,j} + B\sum_{ij \in E_1 }\sum_{\substack{uv \notin E_2 \\ u \neq v }} x_{u,i}x_{v,j}.
\end{split}
\end{equation}

\end{widetext}
In Calude \cite{calude2017qubo} a different methods for constructing efficient QUBO formulations for the Graph Isomorphism Problem. The objective for the formulation is given as follows:
\begin{equation}
F(x) = H(x) + \sum_{ij \in E_1} P_{i,j}(x),
\end{equation}
where 
\begin{equation}
    H(x) = \sum_{0\le i < n}\left( 1 - \sum_{0\le i' <n} x_{i,i'}\right)^2 + \sum_{0\le i' < n}\left( 1 - \sum_{0\le i <n} x_{i,i'}\right)^2,
\end{equation}
and
\begin{equation}
   \sum_{ij \in E_1} P_{i,j}(x) = \sum_{0\le i' < n}\left( x_{i,i'}\sum_{0\le j' < n} x_{j,j'} \left( 1-e_{i', j'}\right)\right),
\end{equation}
$e_{i,j} = 1$ if $ij \in E_2$ and $e_{i,j} = 0$ if
$ij \notin E_2$. This direct formulation of the quadratic function is done such that a mapping $f$ can be decoded from that values of the variables in the quadratic program using a partial function $D$, such that $D: \mathbb{Z}_{2}^{n^2} \mapsto \mathcal{F} $, where $\mathcal{F}$ is the set of all bijections between $V_1$ and $V_2$. The domain of $D$ consists of all vectors $\textbf{x} \in \mathbb{Z}_{2}^{n^2}$. Calude et. al. ensures that the function $H(x) = 0$ if and only if $D$ is defined (and bijective). The component $\sum_{ij \in E_1} P_{i,j}(x) = 0$ if and only if the mapping $f = D(\textbf{x})$ is edge invariant. Therefore the function $F(\textbf{x}) = 0$, for all $\textbf{x} \in \mathbb{Z}_{2}^{n^2}$ if and only if $f = D(\textbf{x})$ is an isomorphism.

\section{\label{sec:level4}Methods}
The methodology for each experiment is different but was done on a common dataset unless stated otherwise. The data generator uses the \textit{NetworkX} library and generates pairs of isomorphic and non-isomorphic connected graphs. The generated graphs can be found in the GitHub repository mentioned in the Data Availability section. For graphs having four nodes, the edge count remains in between four and five. On the other hand, for graphs having five nodes, all the graphs have five edges. Since the size of the QUBO matrix for a connected pair of graphs having $N$ nodes is $N \times N$, thereby utilizing $N^2$ qubits, the problem sizes considered for this work were kept small for NISQ era hardware simulations and computational constraints.  The QUBO matrix was generated using GSGMorph \cite{gsgmorph} from which the quadratic programs were obtained. For each set of problem instances, the offset value for all experiments is determined empirically such that the problem yields an optimal function value of $\theta$ when the graphs are isomorphic, and the optimal function would raise a positive value when the graphs are non-isomorphic. This offset tuning is necessary because QUBO encodings can introduce constant energy shifts; setting the baseline ensures that isomorphic cases yield exactly zero cost.

\subsection{Numerical Setup for QAOA}
The Quantum Approximate Optimization Algorithm (QAOA) was implemented in Qiskit using the \texttt{QAOA} class. The quadratic program, formulated following the approach of Calude et al.~\cite{calude2017qubo}, was first expressed in IBM’s \texttt{Docplex} framework, which enabled its direct translation into the corresponding problem Hamiltonian. This Hamiltonian was then supplied to the \texttt{QAOA} class through the minimum eigensolver interface, which evaluates the minimum eigenvalue associated with the encoded cost function. We restrict to $p = 1$ to remain in the shallow-circuit regime typical of NISQ demonstrations, balancing fidelity with feasibility. Here, $p$ denotes the number of QAOA layers, with each layer consisting of one application of the cost Hamiltonian and one application of the mixer Hamiltonian. We restrict to $p=1$, corresponding to the shallowest nontrivial QAOA circuit with two variational parameters, in order to remain within the NISQ-relevant shallow-depth regime.

\subsection{Numerical Setup for VQE}
The Variational Quantum Eigensolver was implemented using the \texttt{VQE} class in Qiskit. The quadratic program was first expressed in IBM’s \texttt{Docplex} optimization framework, which served as the interface for constructing the corresponding problem Hamiltonian. For the variational ansatz, we employed the \texttt{TwoLocal} circuit consisting of alternating layers of single-qubit $R_Y$ rotations and controlled-$Z$ entangling gates, with five repetitions and a linear entanglement topology. The optimization of circuit parameters was performed using the Simultaneous Perturbation Stochastic Approximation (SPSA) optimizer, with a maximum of 600 iterations, as implemented within Qiskit’s \texttt{MinimumEigenOptimizer} class.

\begin{figure*}
\begin{center}
    \begin{tabular}{cc}
        \subfloat[]{\includegraphics[width = 3.2in]{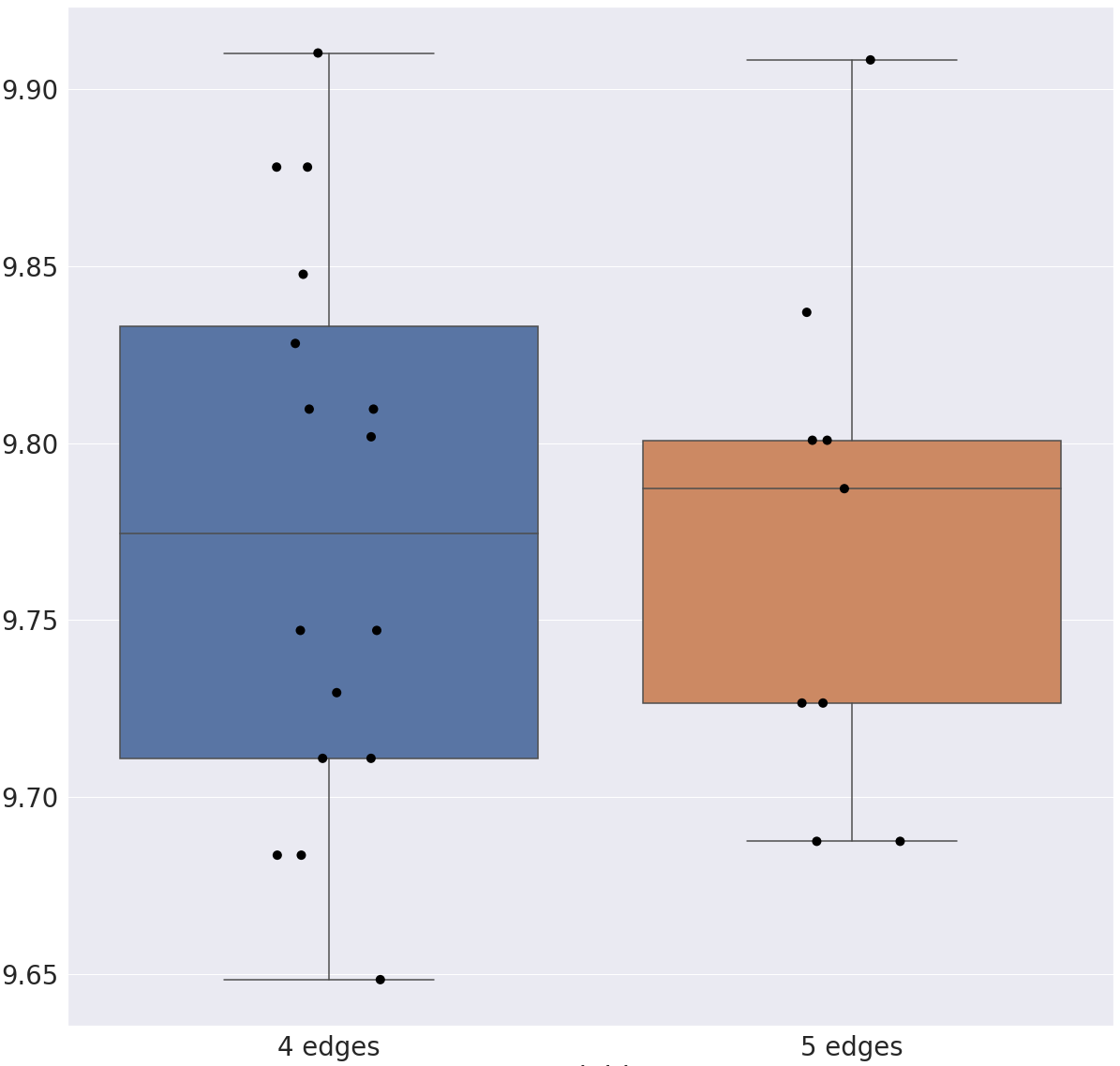}} &
        \subfloat[]{\includegraphics[width = 3.2in]{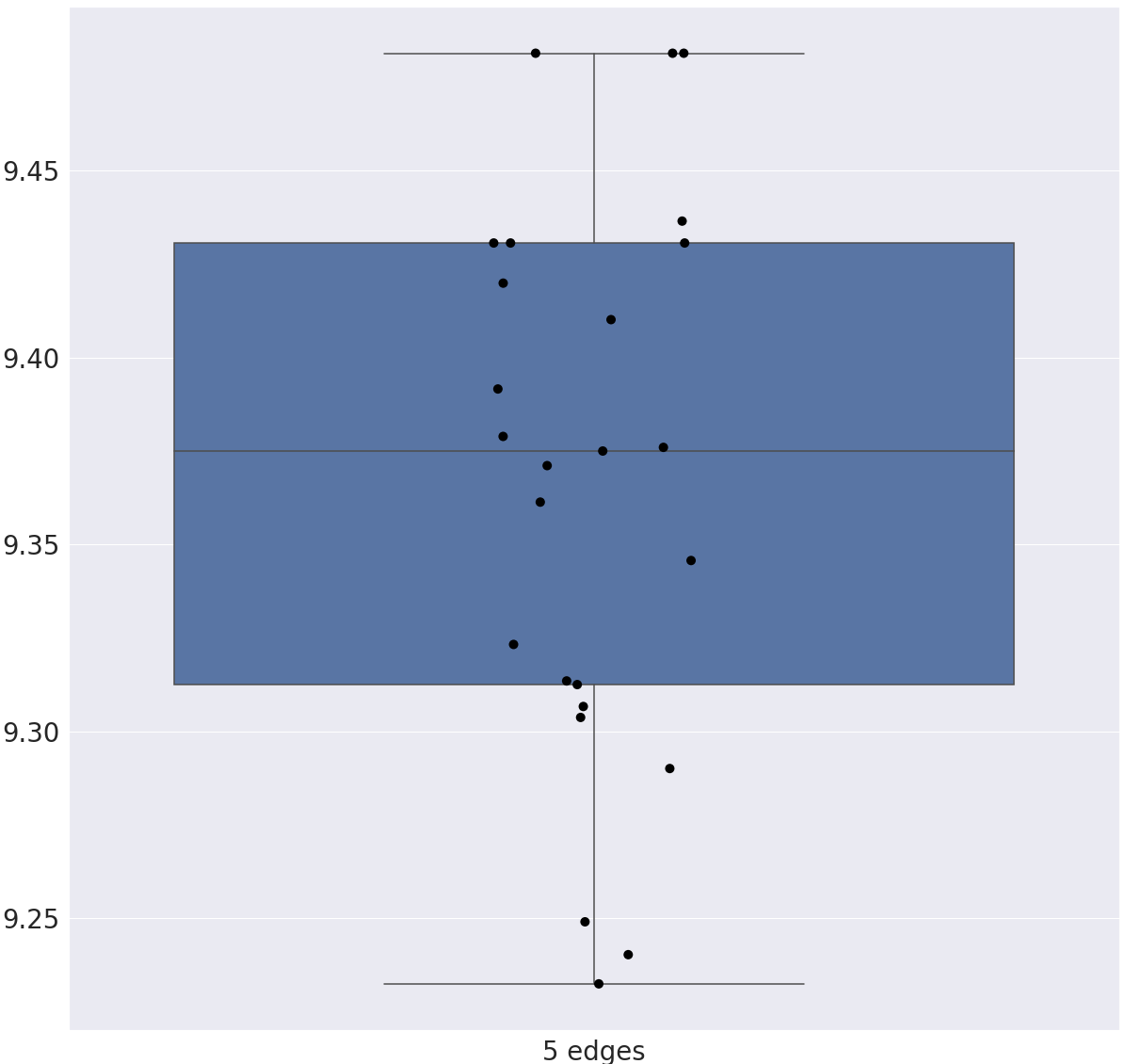}} 
  \end{tabular}
  \caption{The energy values of the QAOA for which the resulting values of their corresponding quadratic values yield an optimal function value of zero. (a) Shows the plot of the energy values for which the values of the quadratic program for the Graph Isomorphism Problem  for a four node graph yields a function value of zero. (b) Shows the plot of the energy values for which the values of the quadratic program for the Graph Isomorphism Problem for a five node graph yields a function value of zero. The \textit{boxplot} shows clearly the distribution of values, the medians and standard deviations for the absolute energy values.}
  \label{fig:qaoa-boxplots}
\end{center}
\end{figure*}

\subsection{Limitations}

Our experiments were restricted to graphs of order four and five due to both hardware and algorithmic constraints. In the QUBO formulation, the number of logical qubits scales as $O(n^2)$ with the number of vertices $n$, making even moderately larger instances infeasible on current NISQ devices with limited qubit counts, connectivity, and coherence times. Classical simulation is likewise hindered by the exponential growth of the Hilbert space. Further limitations arose from algorithmic overhead. The VQE, even with efficient optimizers such as SPSA, is challenging to scale to larger graphs, and remains resource-intensive. For these reasons, we confined our study to small graphs in order to establish proof-of-principle results. This choice enabled us to demonstrate energy clustering for isomorphic graph families under controlled conditions, while leaving the extension to larger and more complex graphs to future work. 

However, we emphasize that the presence of energy clustering does not imply the existence of a strict energy threshold capable of separating isomorphic from non-isomorphic graph pairs. In practice, we observe substantial overlap in the low-energy spectra of isomorphic and non-isomorphic instances, particularly at shallow circuit depth. This overlap is a consequence of the highly degenerate structure of the isomorphism Hamiltonian and the fact that many low-energy states correspond to infeasible assignments that violate permutation constraints. Accordingly, when two data points in Fig.~\ref{fig:qaoa-boxplots} share the same or nearly the same energy, this indicates convergence to energetically similar configurations, not a positive identification of graph isomorphism. To assess how fundamentally limited energy-based discrimination is within our variational framework, we further employed a suite of classical machine learning models trained on quantum-derived features, including final variational energy, optimization depth, and energy descent characteristics. These models—ranging from linear classifiers to expressive ensemble methods—represent a strong post-processing baseline for extracting structure from the variational outputs.

Despite this, classification performance remains modest, with no model achieving reliable separation between isomorphic and non-isomorphic graph pairs. Importantly, the failure of these models cannot be attributed to insufficient classifier expressivity, but rather indicates that the variational signal itself lacks sufficient discriminatory information. This suggests that the observed overlap in low-energy spectra reflects an intrinsic limitation of shallow variational encodings of the isomorphism Hamiltonian, rather than a shortcoming of any particular post-processing strategy. Progress in compact QUBO encodings, problem-tailored ansätze, and error mitigation techniques will be necessary to overcome these current limitations.

\section{Statistical Post-Processing Framework}

The output of a Variational Quantum Algorithm (VQA) for the Graph Isomorphism problem is a stochastic energy expectation value, $\langle H_C \rangle$, rather than a binary decision variable. In an ideal, noise-free environment, isomorphic pairs would converge to a ground state energy of exactly zero, while non-isomorphic pairs would yield a strictly positive residual. However, in the NISQ era, this separation is obscured by stochastic noise, spectral crowding, and imperfect optimization convergence. 

To rigorously distinguish the quantum signal from these noise sources, raw energy values are insufficient. We therefore employ a multi-tiered analysis suite—ranging from dynamic thresholding to manifold learning—to map the decision boundaries of the quantum solver and quantify the distinguishability of the distributions.

\section{Results and discussion}

\subsection{QAOA Energy Landscapes}
The energies calculated correspond to the minimum eigenvalues of the problem Hamiltonian derived via the \texttt{MinimumEigenOptimizer}. The distributions of these energies for graph orders $N=4$ and $N=5$ are presented in \cref{fig:qaoa-boxplots}. 

Consistent with Ising-based formulations of graph isomorphism~\cite{2014P}, we observe that isomorphic graph pairs tend to concentrate around lower variational energy values for small system sizes. For $N=4$, this separation is visually apparent. However, as the number of vertices increases, the ability to discriminate graph classes based solely on the lowest observed energies rapidly degrades. In particular, for $N=5$, we find that the lowest-energy samples obtained by shallow QAOA frequently correspond to infeasible assignments that violate the permutation (bijection) constraints, despite achieving near-optimal objective values.

This phenomenon is a direct consequence of the geometry of the feasible set and the structure of penalty-based encodings. The set of valid permutation matrices constitutes only $N!$ configurations within a Hilbert space of dimension $2^{N^2}$, yielding a vanishingly small feasible fraction
\begin{equation}
\frac{N!}{2^{N^2}} \xrightarrow[N\to\infty]{} 0,
\end{equation}
so that low-energy states are overwhelmingly dominated by infeasible superpositions unless the variational ansatz explicitly enforces permutation structure. Penalty Hamiltonians introduce large degenerate manifolds of near-feasible configurations, leading shallow-depth variational optimization to favor smooth constraint violations over discrete feasibility.

While Szegedy~\cite{szegedy2019qaoa} showed that QAOA expectation values are invariant under graph isomorphisms at fixed depth, this invariance does not imply completeness for graph discrimination, nor does it guarantee that the lowest-energy states encode valid isomorphisms. At shallow depth, the locality horizon of QAOA limits its sensitivity to global bijection constraints, causing distinct non-isomorphic graphs to exhibit overlapping low-energy spectra once feasibility is taken into account. Consequently, discrimination based on minimum energy alone becomes unreliable beyond the smallest instances, even when apparent clustering persists at higher energy scales.

Importantly, the final energy value represents only the endpoint of a variational optimization trajectory. Two graph instances may converge to similar final energies while exhibiting qualitatively different optimization dynamics, such as convergence speed, oscillatory behavior, or energy drop magnitude across iterations. This observation suggests that additional information encoded in the optimization process itself may carry discriminative structure that is not captured by the final energy alone.

To operationalize this insight, we construct a low-dimensional quantum-derived feature vector
\begin{equation}
\mathbf{x} = \bigl[ N,\; E_{\mathrm{final}},\; \text{steps},\; \Delta E \bigr],
\end{equation}
where $N$ is the number of vertices, $E_{\mathrm{final}}$ is the terminal variational energy, $\text{steps}$ denotes the number of optimizer iterations, and $\Delta E$ is the total energy decrease over the optimization trajectory. We then benchmark a suite of standard classical classifiers on this feature set to assess whether trajectory-level information improves discrimination between isomorphic and non-isomorphic graph pairs.

\begin{figure}[h]
    \centering
    \includegraphics[width=\linewidth]{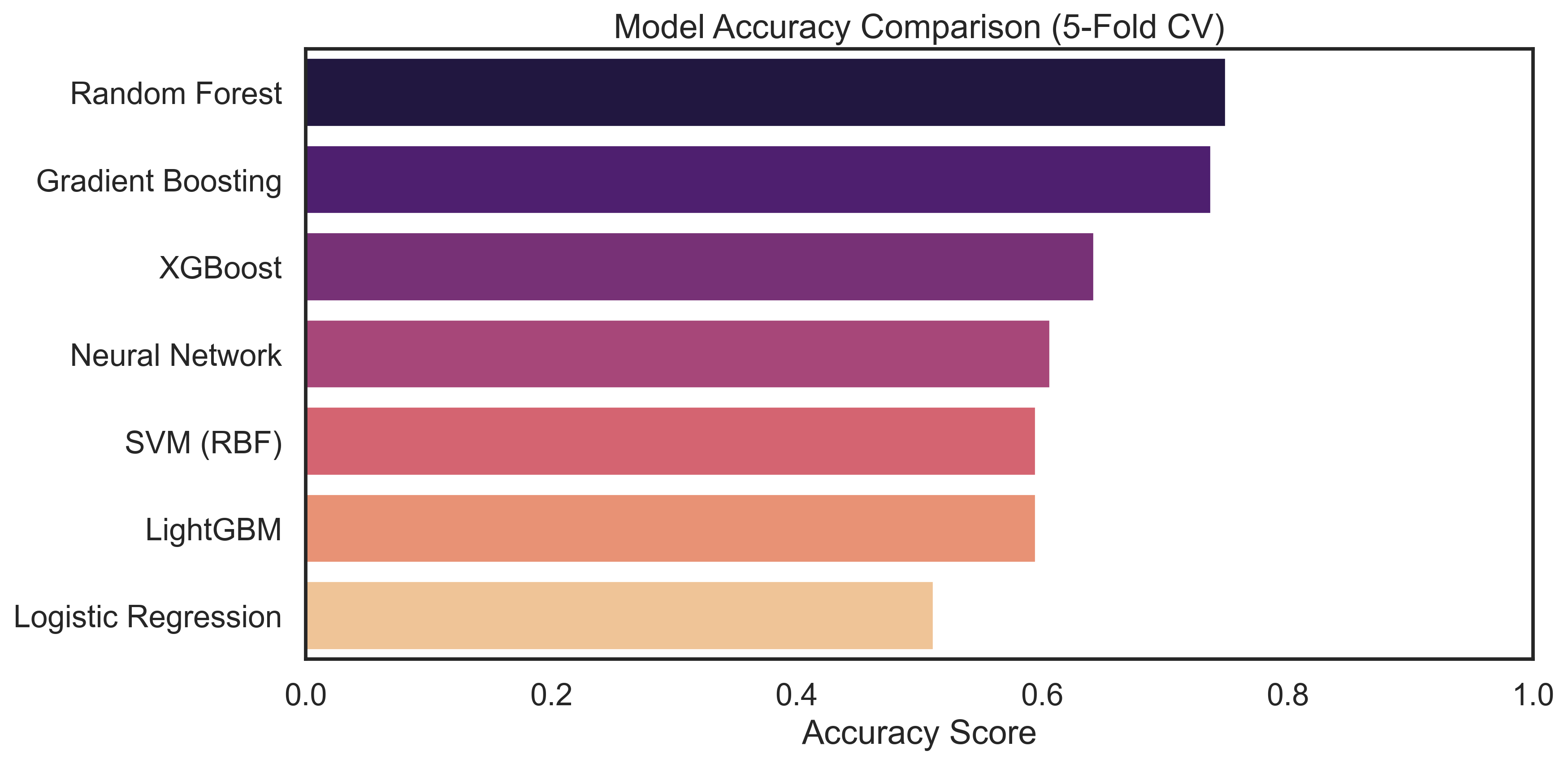}
    \caption{\textbf{Model Performance Comparison.} Tree-based ensemble methods (e.g., Random Forest and Gradient Boosting) consistently outperform linear classifiers, indicating that the decision boundary induced by the quantum-derived feature space is inherently non-linear.}
    \label{fig:model_comparison}
\end{figure}

\begin{figure}[h]
    \centering
    \includegraphics[width=\linewidth]{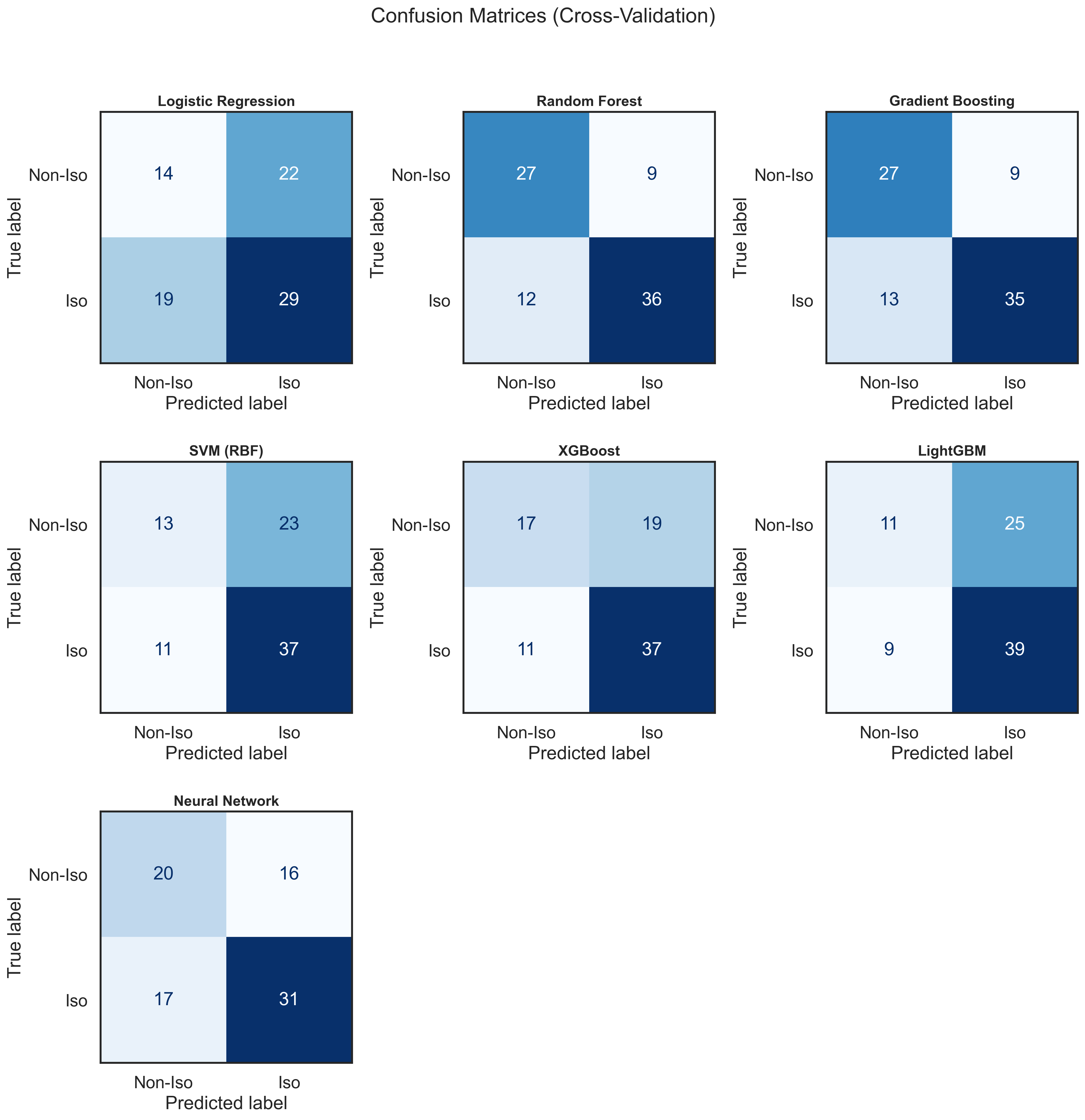}
    \caption{\textbf{Confusion Matrix Grid.} Confusion matrices for each classifier, illustrating the relative prevalence of Type~I (false positive) and Type~II (false negative) errors.}
    \label{fig:confusion_matrices}
\end{figure}

Specifically, we evaluate seven models: Logistic Regression, SVM with RBF kernel, Random Forest, Gradient Boosting, XGBoost, LightGBM, and a Multi-Layer Perceptron (MLP). Figure~\ref{fig:model_comparison} summarizes their classification performance, while the corresponding confusion matrices are shown in Figure~\ref{fig:confusion_matrices} to highlight the dominant error modes.

Quantitative performance metrics for all models are reported in Table~\ref{tab:model_performance}. Consistent with the trends observed in Figure~\ref{fig:model_comparison}, ensemble-based methods achieve the highest accuracy and AUC, with Random Forest performing best overall. In contrast, linear models perform near chance level, reinforcing the conclusion that discrimination relies on non-linear interactions among trajectory-level features rather than simple thresholding on final energy.

\begin{table}[h]
\centering
\caption{Model Performance Ranking}
\label{tab:model_performance}
\begin{tabular}{lcc}
\hline
\textbf{Model} & \textbf{Accuracy} & \textbf{AUC} \\
\hline
Random Forest        & 0.7500 & 0.7917 \\
Gradient Boosting    & 0.7381 & 0.7905 \\
XGBoost              & 0.6429 & 0.7196 \\
Neural Network       & 0.6071 & 0.6424 \\
SVM (RBF)            & 0.5952 & 0.6094 \\
LightGBM             & 0.5952 & 0.6745 \\
Logistic Regression  & 0.5119 & 0.5966 \\
\hline
\end{tabular}
\end{table}

\begin{figure*}
\centering
\begin{tabular}{cc}
    \subfloat[]{\includegraphics[width = 3.2in]{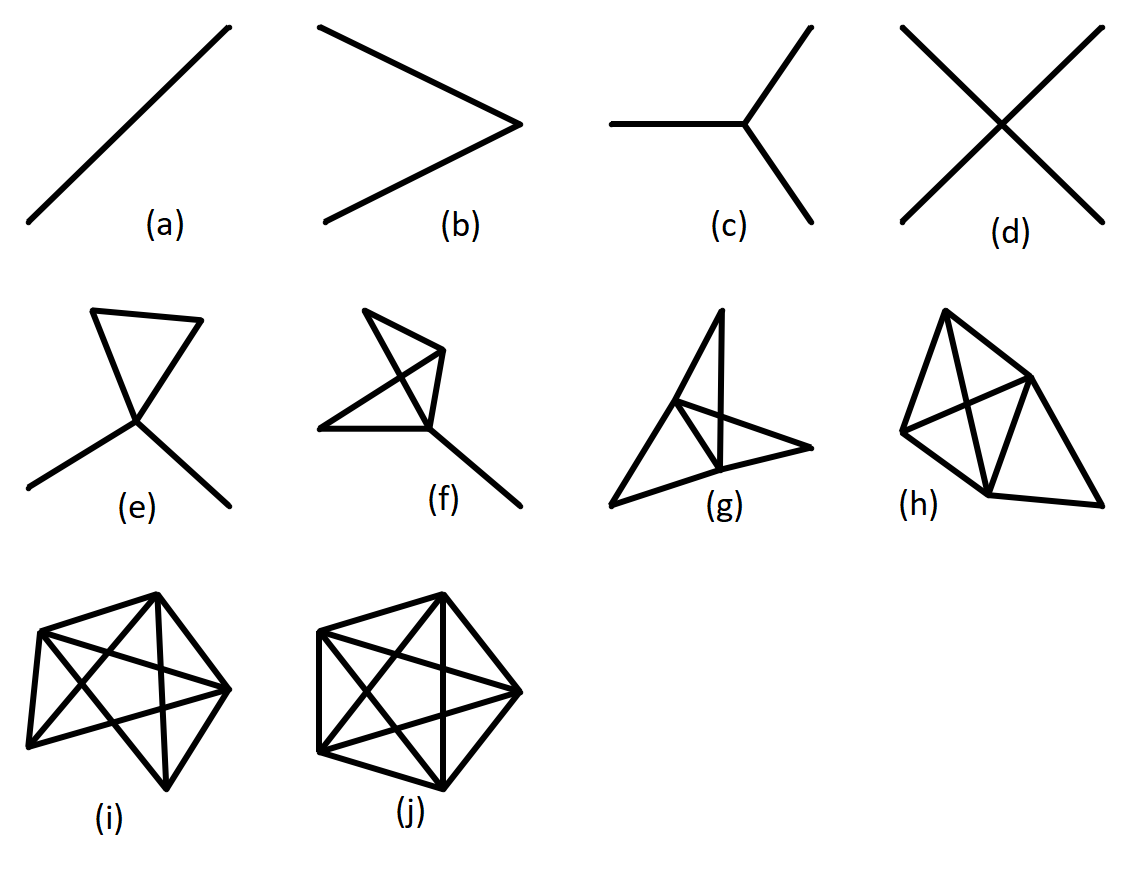}} &
    \subfloat[]{\includegraphics[width = 3.2in]{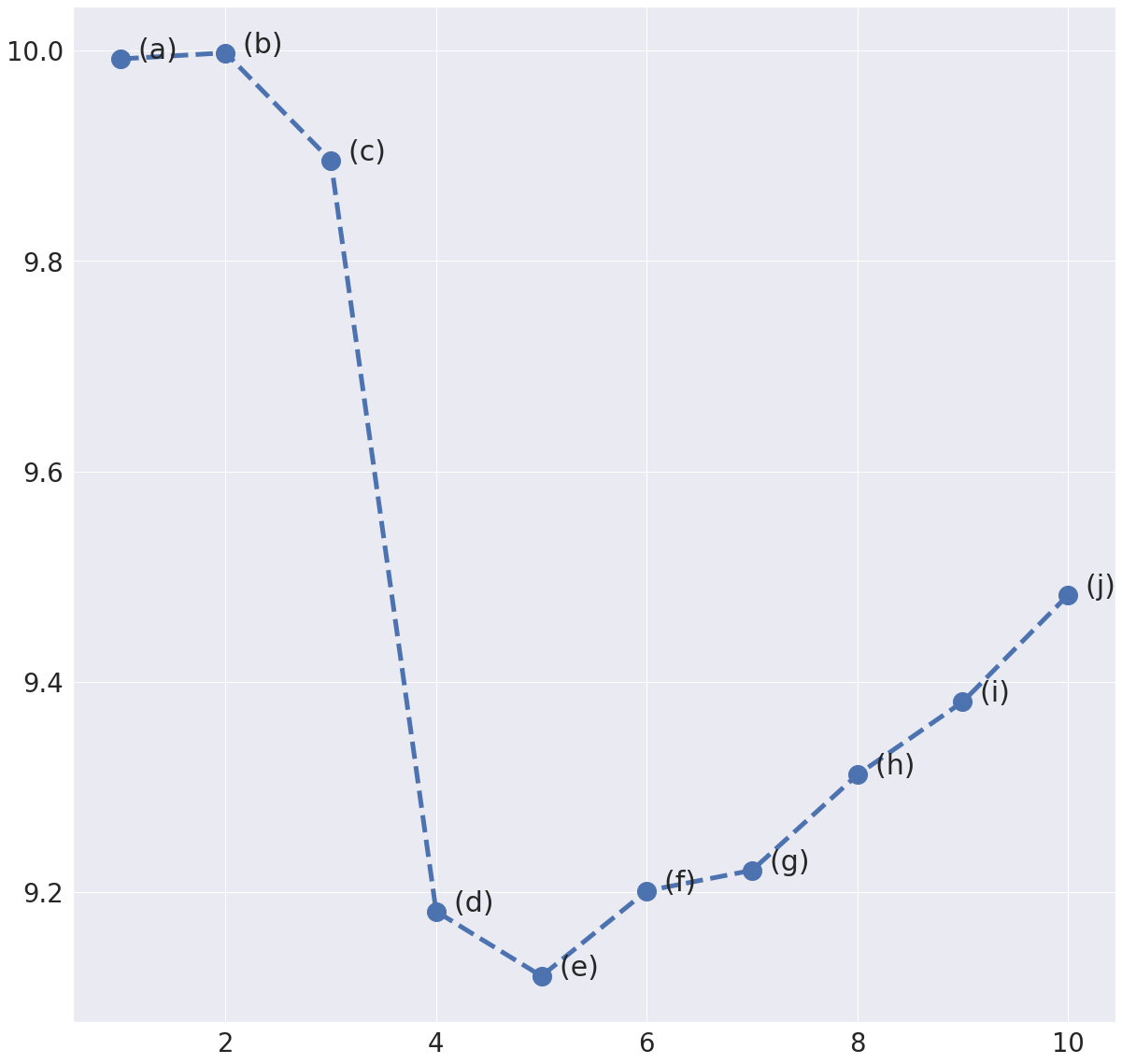}} 
\end{tabular}
\caption{\textbf{Energy response under controlled graph reduction.}
(a) A sequence of graphs obtained by systematic edge removal from the five-node complete graph. For each reduced graph, a copy of the same graph is used, yielding strictly isomorphic pairs.
(b) Final variational energies obtained from QAOA for each isomorphic pair. As edges are removed, the Hamiltonian energy shifts accordingly, while remaining tightly clustered near the minimum for each isomorphism class. This experiment isolates the dependence of the energy landscape on graph structure rather than graph labeling.}
\label{fig:edge-reduction}
\end{figure*}

\subsection{VQE as a variational probe of the isomorphism Hamiltonian}

The QAOA results discussed above characterize the behavior of a shallow-depth variational circuit applied to the graph isomorphism Hamiltonian. To further probe the structure of the same Hamiltonian, we additionally study the Variational Quantum Eigensolver (VQE), which employs a deeper and more expressive ansatz together with a generic variational optimization strategy.

For the small instances considered here, VQE reliably converges to the ground-state energy of the encoded Hamiltonian. In particular, for isomorphic graph pairs, the optimized objective value approaches zero, while non-isomorphic pairs converge to strictly positive values, consistent with the construction of the QUBO formulation. Importantly, this agreement should not be interpreted as evidence of efficient graph discrimination: rather, it confirms that the Hamiltonian correctly encodes isomorphism at the level of its ground state when the variational search is sufficiently expressive.

A notable distinction between QAOA and VQE lies in the scale and structure of the energies encountered. Due to its deeper ansatz and larger parameter space, VQE explores a much broader region of Hilbert space, leading to energy values that are typically orders of magnitude larger than those observed in shallow QAOA runs. This difference reflects improved approximation of the true ground state, rather than a fundamentally different encoding of graph structure.

Despite these differences, both algorithms exhibit a common qualitative trend: energies associated with isomorphic graph families cluster tightly, while non-isomorphic instances populate higher-energy regions of the spectrum. This clustering behavior therefore arises from the structure of the isomorphism Hamiltonian itself, rather than from a specific choice of ansatz or optimizer.

\begin{figure*}
\centering
\begin{tabular}{cc}
    \subfloat[]{\includegraphics[width = 3.5in]{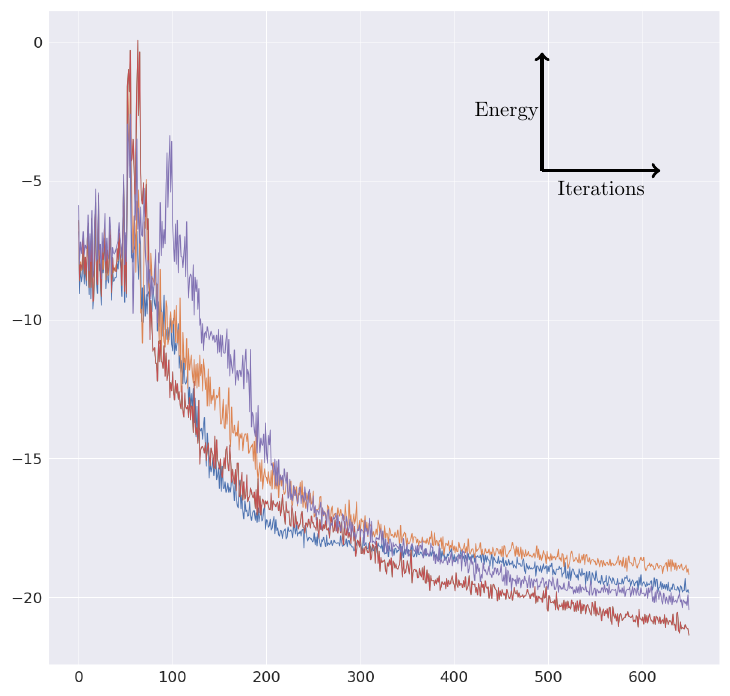}} &
    \subfloat[]{\includegraphics[width = 3.5in]{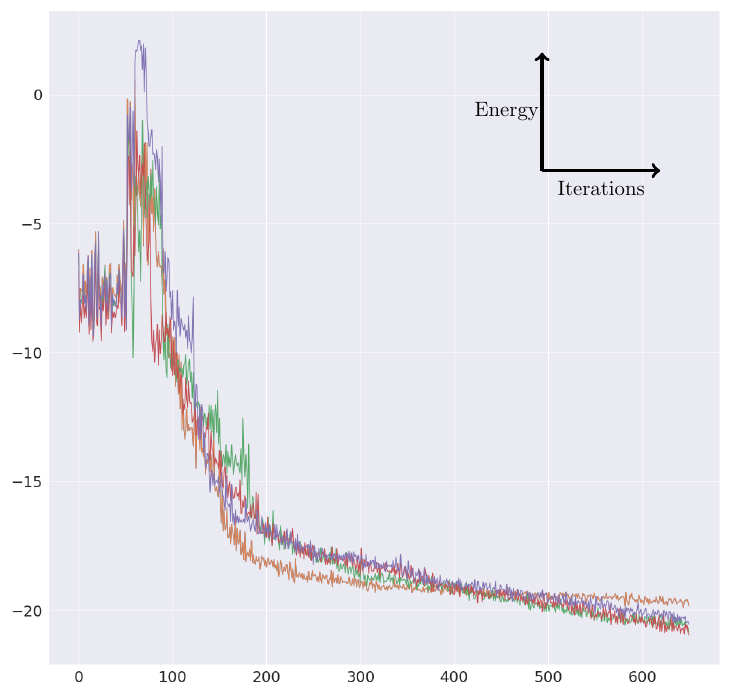}} 
\end{tabular}
\caption{\textbf{VQE convergence behavior for four-node graph pairs.}
Shown are variational energies as a function of optimization iteration for multiple independent VQE runs applied to the same four-node graph pair. Each curve corresponds to a distinct random initialization of the variational parameters (distinguished by both color and line style to aid accessibility). Panels (a) and (b) correspond to two representative graph classes considered in this work (specified in the main text). After approximately 400 iterations, all runs converge to a narrow band of final energies, illustrating \emph{energy clustering}, i.e., low variance in converged energies across runs. This clustering indicates reproducible convergence to low-energy regions of the Hamiltonian landscape, but does not by itself imply discrimination between isomorphic and non-isomorphic graph instances.}
\label{fig:vqe-convergence}
\end{figure*}

The increased expressivity of VQE comes at a substantial computational cost. In our implementation, VQE employed a \texttt{TwoLocal} ansatz optimized using the SPSA algorithm over up to 600 iterations, requiring repeated circuit evaluations at each step. Even for four-node graphs, this resulted in significantly longer runtimes and higher resource consumption compared to QAOA. Figure~\ref{fig:vqe-convergence} illustrates representative convergence traces for multiple VQE runs on four-node graph pairs, showing that energy values stabilize only after several hundred optimization steps.

To disentangle algorithmic effects from properties intrinsic to the Hamiltonian encoding, we perform a controlled structural ablation experiment in which graph topology is varied while graph isomorphism is held fixed. Starting from the five-node complete graph, edges are systematically removed to generate a sequence of increasingly sparse graphs, as shown in Fig.~\ref{fig:edge-reduction}(a). For each graph in this sequence, we consider a pair of identical copies, thereby ensuring that all tested instances are strictly isomorphic by construction.

Figure~\ref{fig:edge-reduction}(b) reports the final QAOA energies obtained for each such isomorphic pair. Because graph labeling is held constant up to permutation, any variation in the observed energies cannot be attributed to relabeling artifacts or optimization noise, but instead reflects changes in the underlying Hamiltonian induced by graph structure. The resulting energy profile demonstrates that the variational energy minimum shifts systematically as edges are removed, while remaining tightly clustered within each isomorphism class.

This figure therefore serves as a calibration experiment: it verifies that the Hamiltonian responds sensitively and smoothly to controlled structural changes in the input graph, independent of labeling. Importantly, the observed clustering should not be interpreted as a decision criterion for graph isomorphism. In particular, near-degenerate energies among distinct graph instances (e.g., graphs (d), (f), and (g) in Fig.~\ref{fig:edge-reduction}) directly impact algorithmic performance by preventing reliable discrimination based on variational energy alone. Rather, the clustering illustrates how structural features of the graph are encoded into the energy landscape, providing a baseline against which the limitations of energy-based discrimination discussed later can be understood.

The observation of energy clustering is scientifically relevant because it probes the structure of the variational energy landscape induced by the graph isomorphism Hamiltonian. Clustering demonstrates that, for a fixed graph instance, the Hamiltonian admits broad low-energy basins that are consistently reached by variational optimization, independent of initialization. By \emph{energy clustering}, we mean the empirical observation that repeated variational optimizations of the same Hamiltonian—initialized from different random parameters—converge to final energies whose variance is small compared to the total energy scale of the problem. In this sense, clustering refers to concentration of the final energies within a narrow band, rather than the emergence of well-separated energy gaps between different graph instances. This indicates that the Hamiltonian encoding is stable and reproducible, rather than dominated by optimizer noise or parameter sensitivity.

At the same time, the absence of well-separated clusters across different graph instances highlights a fundamental limitation: while variational algorithms reliably identify low-energy regions, these regions are not uniquely associated with valid isomorphisms or distinct graph classes. Consequently, energy clustering serves both as a diagnostic of Hamiltonian structure and as evidence that energy-based criteria alone are insufficient for scalable graph isomorphism discrimination.

Taken together, these results highlight the complementary roles of QAOA and VQE in analyzing the graph isomorphism Hamiltonian. QAOA, due to its shallow depth and limited expressivity, serves as a lightweight diagnostic tool for probing coarse features of the energy landscape and identifying structural trends. VQE, by contrast, provides a more faithful approximation of the Hamiltonian ground state, at the expense of substantially increased computational resources.

Crucially, neither approach circumvents the fundamental difficulty of enforcing global permutation constraints within a variational framework. As demonstrated elsewhere in this work, low variational energy does not reliably imply feasibility of the decoded permutation, particularly at shallow depth. Consequently, while energy clustering is a robust feature of the Hamiltonian encoding, it should be interpreted as a property of the variational energy landscape rather than as a complete or scalable solution to the graph isomorphism problem.

\section{\label{sec:level6}Conclusion and Future Work}

In this work, we examined variational quantum approaches to the graph isomorphism problem through a detailed study of the energy landscapes induced by standard QUBO encodings. Using both QAOA and VQE, we analyzed small graph instances and demonstrated that variational optimization reliably produces structured, reproducible low-energy behavior for isomorphic graph pairs. Controlled ablation experiments further confirmed that these energy responses are driven by graph topology rather than labeling artifacts, validating the fidelity of the Hamiltonian encoding in the variational setting.

At the same time, our results make clear that low variational energy alone does not constitute a reliable or scalable criterion for graph isomorphism. We explicitly observed substantial overlap between the low-energy spectra of isomorphic and non-isomorphic instances, as well as frequent convergence to energetically favorable but infeasible configurations that violate permutation constraints. These effects persist even when incorporating trajectory-level information and classical post-processing, underscoring intrinsic limitations of energy-based discrimination at shallow circuit depth.

Taken together, these findings position variational algorithms not as solvers for the graph isomorphism problem, but as diagnostic tools for probing the structure and limitations of Hamiltonian encodings in the NISQ regime. By characterizing both the strengths and failure modes of QAOA and VQE in this context, this work provides a principled baseline for future efforts that seek to incorporate feasibility-preserving mixers, constraint-aware ansätze, or hybrid classical-quantum strategies to move beyond purely energetic heuristics.

A further extension of the investigation pursued in this work involves expanding the idea of energies clustering for families of isomorphic graphs and running the formulation on current devices with greater qubit number. Another natural next step for this line of research is to scale the variational formulations to graphs with 6–8 nodes using more efficient encodings that reduce qubit overhead. While the quadratic growth of qubits in the QUBO formulation presents a bottleneck, recent work on compact encodings and symmetry exploitation could make moderately larger instances tractable. In parallel, testing these formulations on real quantum hardware such as IBM’s superconducting processors or Google’s Sycamore device would provide valuable insights into the effects of noise, decoherence, and limited circuit depth. Employing error mitigation strategies, including zero-noise extrapolation, measurement error correction, and dynamical decoupling, would allow us to benchmark the robustness of the observed energy clustering under realistic device conditions and thereby bridge the gap between proof-of-principle simulations and experimental feasibility.

\begin{acknowledgments}
We thank Dr. Farhan T. Chowdhury at the Ohio State University for a thorough evaluation of our initial draft and advising on key changes that we have implemented in the final manuscript. TC and SIM would like to thank Dr. Omar Shehab from IBM Quantum for insightful discussions during the sessions of Banglay Quantum Computing, and Dr. Amit Saha for his unrelenting support. AK would like to thank Dr. \"{O}zlem Salehi for the fruitful discussion regarding the QUBO formulation, and acknowledges support from the National Science Centre (NCN), Poland, under project number 2019/33/B/ST6/02011.
\end{acknowledgments}

\section*{Data Availability Statement}

The data and codes that support the findings of this study are openly available at 
\href{https://github.com/LegacYFTw/Graph-Isomorphism}{https://github.com/LegacYFTw/Graph-Isomorphism}.

% \nocite{*}
% \bibliography{aipsamp}% Produces the bibliography via BibTeX.

\begin{thebibliography}{43}%
\makeatletter
\providecommand \@ifxundefined [1]{%
 \@ifx{#1\undefined}
}%
\providecommand \@ifnum [1]{%
 \ifnum #1\expandafter \@firstoftwo
 \else \expandafter \@secondoftwo
 \fi
}%
\providecommand \@ifx [1]{%
 \ifx #1\expandafter \@firstoftwo
 \else \expandafter \@secondoftwo
 \fi
}%
\providecommand \natexlab [1]{#1}%
\providecommand \enquote  [1]{``#1''}%
\providecommand \bibnamefont  [1]{#1}%
\providecommand \bibfnamefont [1]{#1}%
\providecommand \citenamefont [1]{#1}%
\providecommand \href@noop [0]{\@secondoftwo}%
\providecommand \href [0]{\begingroup \@sanitize@url \@href}%
\providecommand \@href[1]{\@@startlink{#1}\@@href}%
\providecommand \@@href[1]{\endgroup#1\@@endlink}%
\providecommand \@sanitize@url [0]{\catcode `\\12\catcode `\$12\catcode `\&12\catcode `\#12\catcode `\^12\catcode `\_12\catcode `\%12\relax}%
\providecommand \@@startlink[1]{}%
\providecommand \@@endlink[0]{}%
\providecommand \url  [0]{\begingroup\@sanitize@url \@url }%
\providecommand \@url [1]{\endgroup\@href {#1}{\urlprefix }}%
\providecommand \urlprefix  [0]{URL }%
\providecommand \Eprint [0]{\href }%
\providecommand \doibase [0]{http://dx.doi.org/}%
\providecommand \selectlanguage [0]{\@gobble}%
\providecommand \bibinfo  [0]{\@secondoftwo}%
\providecommand \bibfield  [0]{\@secondoftwo}%
\providecommand \translation [1]{[#1]}%
\providecommand \BibitemOpen [0]{}%
\providecommand \bibitemStop [0]{}%
\providecommand \bibitemNoStop [0]{.\EOS\space}%
\providecommand \EOS [0]{\spacefactor3000\relax}%
\providecommand \BibitemShut  [1]{\csname bibitem#1\endcsname}%
\let\auto@bib@innerbib\@empty
%</preamble>
\bibitem [{\citenamefont {Babai}(2016)}]{babai2016graph}%
  \BibitemOpen
  \bibfield  {author} {\bibinfo {author} {\bibfnamefont {L.}~\bibnamefont {Babai}},\ }\bibfield  {title} {\enquote {\bibinfo {title} {Graph isomorphism in quasipolynomial time},}\ }in\ \href {https://doi.org/10.1145/2897518.2897542} {\emph {\bibinfo {booktitle} {Proceedings of the forty-eighth annual ACM symposium on Theory of Computing}}}\ (\bibinfo {year} {2016})\ pp.\ \bibinfo {pages} {684--697}\BibitemShut {NoStop}%
\bibitem [{\citenamefont {Doga}\ \emph {et~al.}(2024)\citenamefont {Doga}, \citenamefont {Raubenolt}, \citenamefont {Cumbo}, \citenamefont {Joshi}, \citenamefont {DiFilippo}, \citenamefont {Qin}, \citenamefont {Blankenberg},\ and\ \citenamefont {Shehab}}]{shehab24}%
  \BibitemOpen
  \bibfield  {author} {\bibinfo {author} {\bibfnamefont {H.}~\bibnamefont {Doga}}, \bibinfo {author} {\bibfnamefont {B.}~\bibnamefont {Raubenolt}}, \bibinfo {author} {\bibfnamefont {F.}~\bibnamefont {Cumbo}}, \bibinfo {author} {\bibfnamefont {J.}~\bibnamefont {Joshi}}, \bibinfo {author} {\bibfnamefont {F.~P.}\ \bibnamefont {DiFilippo}}, \bibinfo {author} {\bibfnamefont {J.}~\bibnamefont {Qin}}, \bibinfo {author} {\bibfnamefont {D.}~\bibnamefont {Blankenberg}}, \ and\ \bibinfo {author} {\bibfnamefont {O.}~\bibnamefont {Shehab}},\ }\bibfield  {title} {\enquote {\bibinfo {title} {A perspective on protein structure prediction using quantum computers},}\ }\href {\doibase 10.1021/acs.jctc.4c00067} {\bibfield  {journal} {\bibinfo  {journal} {J. Chem. Theory Comput.}\ }\textbf {\bibinfo {volume} {20}},\ \bibinfo {pages} {3359--3378} (\bibinfo {year} {2024})}\BibitemShut {NoStop}%
\bibitem [{\citenamefont {Mohtashim}(2025)}]{rinq25}%
  \BibitemOpen
  \bibfield  {author} {\bibinfo {author} {\bibfnamefont {S.~I.}\ \bibnamefont {Mohtashim}},\ }\bibfield  {title} {\enquote {\bibinfo {title} {Rinq: Towards predicting central sites in proteins on current quantum computers},}\ }\href {https://doi.org/10.1016/j.mtquan.2025.100053} {\bibfield  {journal} {\bibinfo  {journal} {Mater. Today Quantum}\ }\textbf {\bibinfo {volume} {7}},\ \bibinfo {pages} {100053} (\bibinfo {year} {2025})}\BibitemShut {NoStop}%
\bibitem [{\citenamefont {Kyriienko}, \citenamefont {Umeano},\ and\ \citenamefont {Holmes}(2025)}]{kyriienko25}%
  \BibitemOpen
  \bibfield  {author} {\bibinfo {author} {\bibfnamefont {O.}~\bibnamefont {Kyriienko}}, \bibinfo {author} {\bibfnamefont {C.}~\bibnamefont {Umeano}}, \ and\ \bibinfo {author} {\bibfnamefont {Z.}~\bibnamefont {Holmes}},\ }\bibfield  {title} {\enquote {\bibinfo {title} {Advantage for discrete variational quantum algorithms in circuit recompilation},}\ }\href {https://arxiv.org/abs/2510.01154} {\bibfield  {journal} {\bibinfo  {journal} {arXiv:2510.01154}\ } (\bibinfo {year} {2025})}\BibitemShut {NoStop}%
\bibitem [{\citenamefont {Cerezo}\ \emph {et~al.}(2025)\citenamefont {Cerezo} \emph {et~al.}}]{cerezo25}%
  \BibitemOpen
  \bibfield  {author} {\bibinfo {author} {\bibfnamefont {M.}~\bibnamefont {Cerezo}} \emph {et~al.},\ }\bibfield  {title} {\enquote {\bibinfo {title} {Does provable absence of barren plateaus imply classical simulability?}}\ }\href {https://doi.org/10.1038/s41467-025-63099-6} {\bibfield  {journal} {\bibinfo  {journal} {Nat. Commun.}\ }\textbf {\bibinfo {volume} {16}},\ \bibinfo {pages} {7907} (\bibinfo {year} {2025})}\BibitemShut {NoStop}%
\bibitem [{\citenamefont {Mohtashim}\ \emph {et~al.}(2025)\citenamefont {Mohtashim}, \citenamefont {Das}, \citenamefont {Chatterjee},\ and\ \citenamefont {Chowdhury}}]{glass25}%
  \BibitemOpen
  \bibfield  {author} {\bibinfo {author} {\bibfnamefont {S.~I.}\ \bibnamefont {Mohtashim}}, \bibinfo {author} {\bibfnamefont {A.}~\bibnamefont {Das}}, \bibinfo {author} {\bibfnamefont {T.}~\bibnamefont {Chatterjee}}, \ and\ \bibinfo {author} {\bibfnamefont {F.~T.}\ \bibnamefont {Chowdhury}},\ }\bibfield  {title} {\enquote {\bibinfo {title} {A near-term quantum simulation of the transverse field ising model hints at glassy dynamics},}\ }\href {https://doi.org/10.1140/epjs/s11734-025-01630-y} {\bibfield  {journal} {\bibinfo  {journal} {Eur. Phys. J. Spec. Top.}\ } (\bibinfo {year} {2025})}\BibitemShut {NoStop}%
\bibitem [{\citenamefont {Mohtashim}, \citenamefont {Mahatabuddin},\ and\ \citenamefont {Jabbar}(2023)}]{10313751}%
  \BibitemOpen
  \bibfield  {author} {\bibinfo {author} {\bibfnamefont {S.~I.}\ \bibnamefont {Mohtashim}}, \bibinfo {author} {\bibfnamefont {S.}~\bibnamefont {Mahatabuddin}}, \ and\ \bibinfo {author} {\bibfnamefont {M.~A.}\ \bibnamefont {Jabbar}},\ }\bibfield  {title} {\enquote {\bibinfo {title} {Harnessing the vqe to simulate quantum chemistry in an undergraduate project: Properties of hydrogen, oxygen and water molecules},}\ }in\ \href {\doibase 10.1109/QCE57702.2023.20329} {\emph {\bibinfo {booktitle} {2023 IEEE International Conference on Quantum Computing and Engineering (QCE)}}},\ Vol.~\bibinfo {volume} {03}\ (\bibinfo {year} {2023})\ pp.\ \bibinfo {pages} {94--100}\BibitemShut {NoStop}%
\bibitem [{\citenamefont {Brokowski}\ \emph {et~al.}(2022)\citenamefont {Brokowski}, \citenamefont {Chowdhury}, \citenamefont {Smith}, \citenamefont {Alvarez}, \citenamefont {Sandeep},\ and\ \citenamefont {Aiello}}]{trev_ieee}%
  \BibitemOpen
  \bibfield  {author} {\bibinfo {author} {\bibfnamefont {T.~J.}\ \bibnamefont {Brokowski}}, \bibinfo {author} {\bibfnamefont {F.~T.}\ \bibnamefont {Chowdhury}}, \bibinfo {author} {\bibfnamefont {L.~D.}\ \bibnamefont {Smith}}, \bibinfo {author} {\bibfnamefont {P.~H.}\ \bibnamefont {Alvarez}}, \bibinfo {author} {\bibfnamefont {S.}~\bibnamefont {Sandeep}}, \ and\ \bibinfo {author} {\bibfnamefont {C.~D.}\ \bibnamefont {Aiello}},\ }\bibfield  {title} {\enquote {\bibinfo {title} {Spin chemistry simulation via hybrid-quantum machine learning},}\ }in\ \href {\doibase 10.1109/QCE53715.2022.00147} {\emph {\bibinfo {booktitle} {2022 IEEE International Conference on Quantum Computing and Engineering (QCE)}}}\ (\bibinfo {year} {2022})\ pp.\ \bibinfo {pages} {867--868}\BibitemShut {NoStop}%
\bibitem [{\citenamefont {Cerezo}\ \emph {et~al.}(2021)\citenamefont {Cerezo}, \citenamefont {Arrasmith}, \citenamefont {Babbush}, \citenamefont {Benjamin}, \citenamefont {Endo}, \citenamefont {Fujii}, \citenamefont {McClean}, \citenamefont {Mitarai}, \citenamefont {Yuan}, \citenamefont {Cincio} \emph {et~al.}}]{2021}%
  \BibitemOpen
  \bibfield  {author} {\bibinfo {author} {\bibfnamefont {M.}~\bibnamefont {Cerezo}}, \bibinfo {author} {\bibfnamefont {A.}~\bibnamefont {Arrasmith}}, \bibinfo {author} {\bibfnamefont {R.}~\bibnamefont {Babbush}}, \bibinfo {author} {\bibfnamefont {S.~C.}\ \bibnamefont {Benjamin}}, \bibinfo {author} {\bibfnamefont {S.}~\bibnamefont {Endo}}, \bibinfo {author} {\bibfnamefont {K.}~\bibnamefont {Fujii}}, \bibinfo {author} {\bibfnamefont {J.~R.}\ \bibnamefont {McClean}}, \bibinfo {author} {\bibfnamefont {K.}~\bibnamefont {Mitarai}}, \bibinfo {author} {\bibfnamefont {X.}~\bibnamefont {Yuan}}, \bibinfo {author} {\bibfnamefont {L.}~\bibnamefont {Cincio}},  \emph {et~al.},\ }\bibfield  {title} {\enquote {\bibinfo {title} {Variational quantum algorithms},}\ }\href {\doibase 10.1038/s42254-021-00348-9} {\bibfield  {journal} {\bibinfo  {journal} {Nat. Rev. Phys}\ }\textbf {\bibinfo {volume} {3}},\ \bibinfo {pages} {625--644} (\bibinfo {year} {2021})}\BibitemShut {NoStop}%
\bibitem [{\citenamefont {Preskill}(2018)}]{Preskill2018quantumcomputingin}%
  \BibitemOpen
  \bibfield  {author} {\bibinfo {author} {\bibfnamefont {J.}~\bibnamefont {Preskill}},\ }\bibfield  {title} {\enquote {\bibinfo {title} {Quantum {C}omputing in the {NISQ} era and beyond},}\ }\href {\doibase 10.22331/q-2018-08-06-79} {\bibfield  {journal} {\bibinfo  {journal} {{Quantum}}\ }\textbf {\bibinfo {volume} {2}},\ \bibinfo {pages} {79} (\bibinfo {year} {2018})}\BibitemShut {NoStop}%
\bibitem [{\citenamefont {Magann}\ \emph {et~al.}(2021)\citenamefont {Magann}, \citenamefont {Arenz}, \citenamefont {Grace}, \citenamefont {Ho}, \citenamefont {Kosut}, \citenamefont {McClean}, \citenamefont {Rabitz},\ and\ \citenamefont {Sarovar}}]{pulse21}%
  \BibitemOpen
  \bibfield  {author} {\bibinfo {author} {\bibfnamefont {A.~B.}\ \bibnamefont {Magann}}, \bibinfo {author} {\bibfnamefont {C.}~\bibnamefont {Arenz}}, \bibinfo {author} {\bibfnamefont {M.~D.}\ \bibnamefont {Grace}}, \bibinfo {author} {\bibfnamefont {T.-S.}\ \bibnamefont {Ho}}, \bibinfo {author} {\bibfnamefont {R.~L.}\ \bibnamefont {Kosut}}, \bibinfo {author} {\bibfnamefont {J.~R.}\ \bibnamefont {McClean}}, \bibinfo {author} {\bibfnamefont {H.~A.}\ \bibnamefont {Rabitz}}, \ and\ \bibinfo {author} {\bibfnamefont {M.}~\bibnamefont {Sarovar}},\ }\bibfield  {title} {\enquote {\bibinfo {title} {From pulses to circuits and back again: A quantum optimal control perspective on variational quantum algorithms},}\ }\href {https://link.aps.org/doi/10.1103/PRXQuantum.2.010101} {\bibfield  {journal} {\bibinfo  {journal} {PRX Quantum}\ }\textbf {\bibinfo {volume} {2}},\ \bibinfo {pages} {010101} (\bibinfo {year} {2021})}\BibitemShut {NoStop}%
\bibitem [{\citenamefont {Peruzzo}\ \emph {et~al.}(2014)\citenamefont {Peruzzo}, \citenamefont {McClean}, \citenamefont {Shadbolt}, \citenamefont {Yung}, \citenamefont {Zhou}, \citenamefont {Love}, \citenamefont {Aspuru-Guzik},\ and\ \citenamefont {O'Brien}}]{peruzzo2014}%
  \BibitemOpen
  \bibfield  {author} {\bibinfo {author} {\bibfnamefont {A.}~\bibnamefont {Peruzzo}}, \bibinfo {author} {\bibfnamefont {J.}~\bibnamefont {McClean}}, \bibinfo {author} {\bibfnamefont {P.}~\bibnamefont {Shadbolt}}, \bibinfo {author} {\bibfnamefont {M.-H.}\ \bibnamefont {Yung}}, \bibinfo {author} {\bibfnamefont {X.-Q.}\ \bibnamefont {Zhou}}, \bibinfo {author} {\bibfnamefont {P.~J.}\ \bibnamefont {Love}}, \bibinfo {author} {\bibfnamefont {A.}~\bibnamefont {Aspuru-Guzik}}, \ and\ \bibinfo {author} {\bibfnamefont {J.~L.}\ \bibnamefont {O'Brien}},\ }\bibfield  {title} {\enquote {\bibinfo {title} {A variational eigenvalue solver on a photonic quantum processor},}\ }\href {\doibase 10.1038/ncomms5213} {\bibfield  {journal} {\bibinfo  {journal} {Nat. Commun.}\ }\textbf {\bibinfo {volume} {5}} (\bibinfo {year} {2014}),\ 10.1038/ncomms5213}\BibitemShut {NoStop}%
\bibitem [{\citenamefont {Farhi}, \citenamefont {Goldstone},\ and\ \citenamefont {Gutmann}(2014)}]{farhi2014quantum}%
  \BibitemOpen
  \bibfield  {author} {\bibinfo {author} {\bibfnamefont {E.}~\bibnamefont {Farhi}}, \bibinfo {author} {\bibfnamefont {J.}~\bibnamefont {Goldstone}}, \ and\ \bibinfo {author} {\bibfnamefont {S.}~\bibnamefont {Gutmann}},\ }\bibfield  {title} {\enquote {\bibinfo {title} {A quantum approximate optimization algorithm},}\ }\href {https://arxiv.org/abs/1411.4028} {\bibfield  {journal} {\bibinfo  {journal} {arXiv:1411.4028}\ } (\bibinfo {year} {2014})}\BibitemShut {NoStop}%
\bibitem [{\citenamefont {Lucas}(2014)}]{10.3389/fphy.2014.00005}%
  \BibitemOpen
  \bibfield  {author} {\bibinfo {author} {\bibfnamefont {A.}~\bibnamefont {Lucas}},\ }\bibfield  {title} {\enquote {\bibinfo {title} {Ising formulations of many np problems},}\ }\href {\doibase 10.3389/fphy.2014.00005} {\bibfield  {journal} {\bibinfo  {journal} {Front. Phys.}\ }\textbf {\bibinfo {volume} {2}},\ \bibinfo {pages} {5} (\bibinfo {year} {2014})}\BibitemShut {NoStop}%
\bibitem [{\citenamefont {Venegas-Andraca}\ \emph {et~al.}(2018)\citenamefont {Venegas-Andraca}, \citenamefont {Cruz-Santos}, \citenamefont {McGeoch},\ and\ \citenamefont {Lanzagorta}}]{2018}%
  \BibitemOpen
  \bibfield  {author} {\bibinfo {author} {\bibfnamefont {S.~E.}\ \bibnamefont {Venegas-Andraca}}, \bibinfo {author} {\bibfnamefont {W.}~\bibnamefont {Cruz-Santos}}, \bibinfo {author} {\bibfnamefont {C.}~\bibnamefont {McGeoch}}, \ and\ \bibinfo {author} {\bibfnamefont {M.}~\bibnamefont {Lanzagorta}},\ }\bibfield  {title} {\enquote {\bibinfo {title} {A cross-disciplinary introduction to quantum annealing-based algorithms},}\ }\href {\doibase 10.1080/00107514.2018.1450720} {\bibfield  {journal} {\bibinfo  {journal} {Contemp. Phys.}\ }\textbf {\bibinfo {volume} {59}},\ \bibinfo {pages} {174--197} (\bibinfo {year} {2018})}\BibitemShut {NoStop}%
\bibitem [{\citenamefont {Santoro}\ and\ \citenamefont {Tosatti}(2006)}]{Santoro_2006}%
  \BibitemOpen
  \bibfield  {author} {\bibinfo {author} {\bibfnamefont {G.~E.}\ \bibnamefont {Santoro}}\ and\ \bibinfo {author} {\bibfnamefont {E.}~\bibnamefont {Tosatti}},\ }\bibfield  {title} {\enquote {\bibinfo {title} {Optimization using quantum mechanics: quantum annealing through adiabatic evolution},}\ }\href {\doibase 10.1088/0305-4470/39/36/r01} {\bibfield  {journal} {\bibinfo  {journal} {J. Phys. A: Math. Gen.}\ }\textbf {\bibinfo {volume} {39}},\ \bibinfo {pages} {R393--R431} (\bibinfo {year} {2006})}\BibitemShut {NoStop}%
\bibitem [{\citenamefont {Lewis}\ and\ \citenamefont {Glover}(2017)}]{lewis2017quadratic}%
  \BibitemOpen
  \bibfield  {author} {\bibinfo {author} {\bibfnamefont {M.}~\bibnamefont {Lewis}}\ and\ \bibinfo {author} {\bibfnamefont {F.}~\bibnamefont {Glover}},\ }\bibfield  {title} {\enquote {\bibinfo {title} {Quadratic unconstrained binary optimization problem preprocessing: Theory and empirical analysis},}\ }\href {https://arxiv.org/abs/1705.09844} {\bibfield  {journal} {\bibinfo  {journal} {arXiv:1705.09844}\ } (\bibinfo {year} {2017})}\BibitemShut {NoStop}%
\bibitem [{\citenamefont {Domino}\ \emph {et~al.}(2021)\citenamefont {Domino}, \citenamefont {Kundu}, \citenamefont {Salehi},\ and\ \citenamefont {Krawiec}}]{domino2021quadratic}%
  \BibitemOpen
  \bibfield  {author} {\bibinfo {author} {\bibfnamefont {K.}~\bibnamefont {Domino}}, \bibinfo {author} {\bibfnamefont {A.}~\bibnamefont {Kundu}}, \bibinfo {author} {\bibfnamefont {{\"O}.}~\bibnamefont {Salehi}}, \ and\ \bibinfo {author} {\bibfnamefont {K.}~\bibnamefont {Krawiec}},\ }\bibfield  {title} {\enquote {\bibinfo {title} {Quadratic and higher-order unconstrained binary optimization of railway dispatching problem for quantum computing},}\ }\href@noop {} {\bibfield  {journal} {\bibinfo  {journal} {arXiv:2107.03234}\ } (\bibinfo {year} {2021})}\BibitemShut {NoStop}%
\bibitem [{\citenamefont {Zick}, \citenamefont {Shehab},\ and\ \citenamefont {French}(2015)}]{2015}%
  \BibitemOpen
  \bibfield  {author} {\bibinfo {author} {\bibfnamefont {K.~M.}\ \bibnamefont {Zick}}, \bibinfo {author} {\bibfnamefont {O.}~\bibnamefont {Shehab}}, \ and\ \bibinfo {author} {\bibfnamefont {M.}~\bibnamefont {French}},\ }\bibfield  {title} {\enquote {\bibinfo {title} {Experimental quantum annealing: case study involving the graph isomorphism problem},}\ }\href {\doibase 10.1038/srep11168} {\bibfield  {journal} {\bibinfo  {journal} {Sci. Rep}\ }\textbf {\bibinfo {volume} {5}} (\bibinfo {year} {2015}),\ 10.1038/srep11168}\BibitemShut {NoStop}%
\bibitem [{\citenamefont {Calude}, \citenamefont {Dinneen},\ and\ \citenamefont {Hua}(2017)}]{calude2017qubo}%
  \BibitemOpen
  \bibfield  {author} {\bibinfo {author} {\bibfnamefont {C.~S.}\ \bibnamefont {Calude}}, \bibinfo {author} {\bibfnamefont {M.~J.}\ \bibnamefont {Dinneen}}, \ and\ \bibinfo {author} {\bibfnamefont {R.}~\bibnamefont {Hua}},\ }\bibfield  {title} {\enquote {\bibinfo {title} {Qubo formulations for the graph isomorphism problem and related problems},}\ }\href@noop {} {\bibfield  {journal} {\bibinfo  {journal} {Theor. Comput. Sci.}\ }\textbf {\bibinfo {volume} {701}},\ \bibinfo {pages} {54--69} (\bibinfo {year} {2017})}\BibitemShut {NoStop}%
\bibitem [{\citenamefont {Farhi}\ \emph {et~al.}(2000)\citenamefont {Farhi}, \citenamefont {Goldstone}, \citenamefont {Gutmann},\ and\ \citenamefont {Sipser}}]{farhi2000quantum}%
  \BibitemOpen
  \bibfield  {author} {\bibinfo {author} {\bibfnamefont {E.}~\bibnamefont {Farhi}}, \bibinfo {author} {\bibfnamefont {J.}~\bibnamefont {Goldstone}}, \bibinfo {author} {\bibfnamefont {S.}~\bibnamefont {Gutmann}}, \ and\ \bibinfo {author} {\bibfnamefont {M.}~\bibnamefont {Sipser}},\ }\bibfield  {title} {\enquote {\bibinfo {title} {Quantum computation by adiabatic evolution},}\ }\href@noop {} {\bibfield  {journal} {\bibinfo  {journal} {arXiv preprint quant-ph/0001106}\ } (\bibinfo {year} {2000})}\BibitemShut {NoStop}%
\bibitem [{\citenamefont {Bauckhage}\ \emph {et~al.}(2017)\citenamefont {Bauckhage}, \citenamefont {Brito}, \citenamefont {Cvejoski}, \citenamefont {Ojeda}, \citenamefont {Sifa},\ and\ \citenamefont {Wrobel}}]{bauckhage2017ising}%
  \BibitemOpen
  \bibfield  {author} {\bibinfo {author} {\bibfnamefont {C.}~\bibnamefont {Bauckhage}}, \bibinfo {author} {\bibfnamefont {E.}~\bibnamefont {Brito}}, \bibinfo {author} {\bibfnamefont {K.}~\bibnamefont {Cvejoski}}, \bibinfo {author} {\bibfnamefont {C.}~\bibnamefont {Ojeda}}, \bibinfo {author} {\bibfnamefont {R.}~\bibnamefont {Sifa}}, \ and\ \bibinfo {author} {\bibfnamefont {S.}~\bibnamefont {Wrobel}},\ }\bibfield  {title} {\enquote {\bibinfo {title} {Ising models for binary clustering via adiabatic quantum computing},}\ }in\ \href@noop {} {\emph {\bibinfo {booktitle} {International Workshop on Energy Minimization Methods in Computer Vision and Pattern Recognition}}}\ (\bibinfo {organization} {Springer},\ \bibinfo {year} {2017})\ pp.\ \bibinfo {pages} {3--17}\BibitemShut {NoStop}%
\bibitem [{\citenamefont {Pudenz}\ and\ \citenamefont {Lidar}(2013)}]{pudenz2013quantum}%
  \BibitemOpen
  \bibfield  {author} {\bibinfo {author} {\bibfnamefont {K.~L.}\ \bibnamefont {Pudenz}}\ and\ \bibinfo {author} {\bibfnamefont {D.~A.}\ \bibnamefont {Lidar}},\ }\bibfield  {title} {\enquote {\bibinfo {title} {Quantum adiabatic machine learning},}\ }\href@noop {} {\bibfield  {journal} {\bibinfo  {journal} {Quantum Inf. Process.}\ }\textbf {\bibinfo {volume} {12}},\ \bibinfo {pages} {2027--2070} (\bibinfo {year} {2013})}\BibitemShut {NoStop}%
\bibitem [{\citenamefont {Salehi}, \citenamefont {Glos},\ and\ \citenamefont {Miszczak}(2021)}]{salehi2021unconstrained}%
  \BibitemOpen
  \bibfield  {author} {\bibinfo {author} {\bibfnamefont {{\"O}.}~\bibnamefont {Salehi}}, \bibinfo {author} {\bibfnamefont {A.}~\bibnamefont {Glos}}, \ and\ \bibinfo {author} {\bibfnamefont {J.~A.}\ \bibnamefont {Miszczak}},\ }\bibfield  {title} {\enquote {\bibinfo {title} {Unconstrained binary models of the travelling salesman problem variants for quantum optimization},}\ }\href@noop {} {\bibfield  {journal} {\bibinfo  {journal} {arXiv:2106.09056}\ } (\bibinfo {year} {2021})}\BibitemShut {NoStop}%
\bibitem [{\citenamefont {Borowski}\ \emph {et~al.}(2020)\citenamefont {Borowski}, \citenamefont {Gora}, \citenamefont {Karnas}, \citenamefont {B{\l}ajda}, \citenamefont {Kr{\'o}l}, \citenamefont {Matyjasek}, \citenamefont {Burczyk}, \citenamefont {Szewczyk},\ and\ \citenamefont {Kutwin}}]{borowski2020new}%
  \BibitemOpen
  \bibfield  {author} {\bibinfo {author} {\bibfnamefont {M.}~\bibnamefont {Borowski}}, \bibinfo {author} {\bibfnamefont {P.}~\bibnamefont {Gora}}, \bibinfo {author} {\bibfnamefont {K.}~\bibnamefont {Karnas}}, \bibinfo {author} {\bibfnamefont {M.}~\bibnamefont {B{\l}ajda}}, \bibinfo {author} {\bibfnamefont {K.}~\bibnamefont {Kr{\'o}l}}, \bibinfo {author} {\bibfnamefont {A.}~\bibnamefont {Matyjasek}}, \bibinfo {author} {\bibfnamefont {D.}~\bibnamefont {Burczyk}}, \bibinfo {author} {\bibfnamefont {M.}~\bibnamefont {Szewczyk}}, \ and\ \bibinfo {author} {\bibfnamefont {M.}~\bibnamefont {Kutwin}},\ }\bibfield  {title} {\enquote {\bibinfo {title} {New hybrid quantum annealing algorithms for solving vehicle routing problem},}\ }in\ \href@noop {} {\emph {\bibinfo {booktitle} {International Conference on Computational Science}}}\ (\bibinfo {organization} {Springer},\ \bibinfo {year} {2020})\ pp.\ \bibinfo {pages} {546--561}\BibitemShut {NoStop}%
\bibitem [{\citenamefont {Ladner}(1975)}]{10.1145/321864.321877}%
  \BibitemOpen
  \bibfield  {author} {\bibinfo {author} {\bibfnamefont {R.~E.}\ \bibnamefont {Ladner}},\ }\bibfield  {title} {\enquote {\bibinfo {title} {On the structure of polynomial time reducibility},}\ }\href {\doibase 10.1145/321864.321877} {\bibfield  {journal} {\bibinfo  {journal} {J. ACM}\ }\textbf {\bibinfo {volume} {22}},\ \bibinfo {pages} {155–171} (\bibinfo {year} {1975})}\BibitemShut {NoStop}%
\bibitem [{\citenamefont {Shehab}\ and\ \citenamefont {Jr.}(2017)}]{shehab2017quantum}%
  \BibitemOpen
  \bibfield  {author} {\bibinfo {author} {\bibfnamefont {O.}~\bibnamefont {Shehab}}\ and\ \bibinfo {author} {\bibfnamefont {S.~J.~L.}\ \bibnamefont {Jr.}},\ }\bibfield  {title} {\enquote {\bibinfo {title} {Quantum fourier sampling is guaranteed to fail to compute automorphism groups of easy graphs},}\ }\href {https://arxiv.org/abs/1705.00760} {\bibfield  {journal} {\bibinfo  {journal} {arXiv:1705.00760}\ } (\bibinfo {year} {2017})}\BibitemShut {NoStop}%
\bibitem [{\citenamefont {Hallgren}\ \emph {et~al.}(2010)\citenamefont {Hallgren}, \citenamefont {Moore}, \citenamefont {R\"{o}tteler}, \citenamefont {Russell},\ and\ \citenamefont {Sen}}]{10.1145/1857914.1857918}%
  \BibitemOpen
  \bibfield  {author} {\bibinfo {author} {\bibfnamefont {S.}~\bibnamefont {Hallgren}}, \bibinfo {author} {\bibfnamefont {C.}~\bibnamefont {Moore}}, \bibinfo {author} {\bibfnamefont {M.}~\bibnamefont {R\"{o}tteler}}, \bibinfo {author} {\bibfnamefont {A.}~\bibnamefont {Russell}}, \ and\ \bibinfo {author} {\bibfnamefont {P.}~\bibnamefont {Sen}},\ }\bibfield  {title} {\enquote {\bibinfo {title} {Limitations of quantum coset states for graph isomorphism},}\ }\href {\doibase 10.1145/1857914.1857918} {\bibfield  {journal} {\bibinfo  {journal} {J. ACM}\ }\textbf {\bibinfo {volume} {57}} (\bibinfo {year} {2010}),\ 10.1145/1857914.1857918}\BibitemShut {NoStop}%
\bibitem [{\citenamefont {Grigni}\ \emph {et~al.}(2001)\citenamefont {Grigni}, \citenamefont {Schulman}, \citenamefont {Vazirani},\ and\ \citenamefont {Vazirani}}]{10.1145/380752.380769}%
  \BibitemOpen
  \bibfield  {author} {\bibinfo {author} {\bibfnamefont {M.}~\bibnamefont {Grigni}}, \bibinfo {author} {\bibfnamefont {L.}~\bibnamefont {Schulman}}, \bibinfo {author} {\bibfnamefont {M.}~\bibnamefont {Vazirani}}, \ and\ \bibinfo {author} {\bibfnamefont {U.}~\bibnamefont {Vazirani}},\ }\bibfield  {title} {\enquote {\bibinfo {title} {Quantum mechanical algorithms for the nonabelian hidden subgroup problem},}\ }in\ \href {\doibase 10.1145/380752.380769} {\emph {\bibinfo {booktitle} {Proceedings of the Thirty-Third Annual ACM Symposium on Theory of Computing}}},\ \bibinfo {series and number} {STOC '01}\ (\bibinfo  {publisher} {Association for Computing Machinery},\ \bibinfo {address} {New York, NY, USA},\ \bibinfo {year} {2001})\ p.\ \bibinfo {pages} {68–74}\BibitemShut {NoStop}%
\bibitem [{\citenamefont {Jozsa}(1998)}]{1998}%
  \BibitemOpen
  \bibfield  {author} {\bibinfo {author} {\bibfnamefont {R.}~\bibnamefont {Jozsa}},\ }\bibfield  {title} {\enquote {\bibinfo {title} {Quantum algorithms and the fourier transform},}\ }\href {\doibase 10.1098/rspa.1998.0163} {\bibfield  {journal} {\bibinfo  {journal} {Proc. R. Soc. Lond., Ser. A, Math. Phys. Eng. Sci.}\ }\textbf {\bibinfo {volume} {454}},\ \bibinfo {pages} {323–337} (\bibinfo {year} {1998})}\BibitemShut {NoStop}%
\bibitem [{\citenamefont {Lancaster}\ and\ \citenamefont {Allen}(2025)}]{lancaster2025simulating}%
  \BibitemOpen
  \bibfield  {author} {\bibinfo {author} {\bibfnamefont {J.~L.}\ \bibnamefont {Lancaster}}\ and\ \bibinfo {author} {\bibfnamefont {D.~B.}\ \bibnamefont {Allen}},\ }\bibfield  {title} {\enquote {\bibinfo {title} {Simulating spin dynamics with quantum computers},}\ }\href {\doibase 10.1119/5.0112717} {\bibfield  {journal} {\bibinfo  {journal} {Am. J. Phys.}\ }\textbf {\bibinfo {volume} {93}},\ \bibinfo {pages} {98--109} (\bibinfo {year} {2025})}\BibitemShut {NoStop}%
\bibitem [{\citenamefont {Alvarez}\ \emph {et~al.}(2024)\citenamefont {Alvarez}, \citenamefont {Chowdhury}, \citenamefont {Smith}, \citenamefont {Brokowski}, \citenamefont {Aiello}, \citenamefont {Kattnig},\ and\ \citenamefont {de~Oliveira}}]{aplQ}%
  \BibitemOpen
  \bibfield  {author} {\bibinfo {author} {\bibfnamefont {P.~H.}\ \bibnamefont {Alvarez}}, \bibinfo {author} {\bibfnamefont {F.~T.}\ \bibnamefont {Chowdhury}}, \bibinfo {author} {\bibfnamefont {L.~D.}\ \bibnamefont {Smith}}, \bibinfo {author} {\bibfnamefont {T.~J.}\ \bibnamefont {Brokowski}}, \bibinfo {author} {\bibfnamefont {C.~D.}\ \bibnamefont {Aiello}}, \bibinfo {author} {\bibfnamefont {D.~R.}\ \bibnamefont {Kattnig}}, \ and\ \bibinfo {author} {\bibfnamefont {M.~C.}\ \bibnamefont {de~Oliveira}},\ }\bibfield  {title} {\enquote {\bibinfo {title} {{Simulating spin biology using a digital quantum computer: Prospects on a near-term quantum hardware emulator}},}\ }\href {\doibase 10.1063/5.0213120} {\bibfield  {journal} {\bibinfo  {journal} {APL Quantum}\ }\textbf {\bibinfo {volume} {1}},\ \bibinfo {pages} {036114} (\bibinfo {year} {2024})}\BibitemShut {NoStop}%
\bibitem [{\citenamefont {Lee}(2023)}]{Lee_2023}%
  \BibitemOpen
  \bibfield  {author} {\bibinfo {author} {\bibfnamefont {Y.}~\bibnamefont {Lee}},\ }\bibfield  {title} {\enquote {\bibinfo {title} {Symmetric trotterization in digital quantum simulation of quantum spin dynamics},}\ }\href {\doibase 10.1007/s40042-023-00722-z} {\bibfield  {journal} {\bibinfo  {journal} {J. Korean Phys. Soc.}\ }\textbf {\bibinfo {volume} {82}},\ \bibinfo {pages} {479--485} (\bibinfo {year} {2023})}\BibitemShut {NoStop}%
\bibitem [{\citenamefont {Farhi}\ \emph {et~al.}(2001)\citenamefont {Farhi}, \citenamefont {Goldstone}, \citenamefont {Gutmann}, \citenamefont {Lapan}, \citenamefont {Lundgren},\ and\ \citenamefont {Preda}}]{farhi2001quantum}%
  \BibitemOpen
  \bibfield  {author} {\bibinfo {author} {\bibfnamefont {E.}~\bibnamefont {Farhi}}, \bibinfo {author} {\bibfnamefont {J.}~\bibnamefont {Goldstone}}, \bibinfo {author} {\bibfnamefont {S.}~\bibnamefont {Gutmann}}, \bibinfo {author} {\bibfnamefont {J.}~\bibnamefont {Lapan}}, \bibinfo {author} {\bibfnamefont {A.}~\bibnamefont {Lundgren}}, \ and\ \bibinfo {author} {\bibfnamefont {D.}~\bibnamefont {Preda}},\ }\bibfield  {title} {\enquote {\bibinfo {title} {A quantum adiabatic evolution algorithm applied to random instances of an np-complete problem},}\ }\href@noop {} {\bibfield  {journal} {\bibinfo  {journal} {Science}\ }\textbf {\bibinfo {volume} {292}},\ \bibinfo {pages} {472--475} (\bibinfo {year} {2001})}\BibitemShut {NoStop}%
\bibitem [{\citenamefont {Majumdar}\ \emph {et~al.}(2021)\citenamefont {Majumdar}, \citenamefont {Madan}, \citenamefont {Bhoumik}, \citenamefont {Vinayagamurthy}, \citenamefont {Raghunathan},\ and\ \citenamefont {Sur-Kolay}}]{majumdar2021optimizing}%
  \BibitemOpen
  \bibfield  {author} {\bibinfo {author} {\bibfnamefont {R.}~\bibnamefont {Majumdar}}, \bibinfo {author} {\bibfnamefont {D.}~\bibnamefont {Madan}}, \bibinfo {author} {\bibfnamefont {D.}~\bibnamefont {Bhoumik}}, \bibinfo {author} {\bibfnamefont {D.}~\bibnamefont {Vinayagamurthy}}, \bibinfo {author} {\bibfnamefont {S.}~\bibnamefont {Raghunathan}}, \ and\ \bibinfo {author} {\bibfnamefont {S.}~\bibnamefont {Sur-Kolay}},\ }\bibfield  {title} {\enquote {\bibinfo {title} {Optimizing ansatz design in qaoa for max-cut},}\ }\href {https://arxiv.org/abs/2106.02812} {\bibfield  {journal} {\bibinfo  {journal} {arXiv:2106.02812}\ } (\bibinfo {year} {2021})}\BibitemShut {NoStop}%
\bibitem [{\citenamefont {Cook}, \citenamefont {Eidenbenz},\ and\ \citenamefont {Bartschi}(2020)}]{2020}%
  \BibitemOpen
  \bibfield  {author} {\bibinfo {author} {\bibfnamefont {J.}~\bibnamefont {Cook}}, \bibinfo {author} {\bibfnamefont {S.}~\bibnamefont {Eidenbenz}}, \ and\ \bibinfo {author} {\bibfnamefont {A.}~\bibnamefont {Bartschi}},\ }\bibfield  {title} {\enquote {\bibinfo {title} {The quantum alternating operator ansatz on maximum k-vertex cover},}\ }in\ \href {\doibase 10.1109/QCE49297.2020.00021} {\emph {\bibinfo {booktitle} {2020 IEEE International Conference on Quantum Computing and Engineering (QCE)}}}\ (\bibinfo {year} {2020})\ pp.\ \bibinfo {pages} {58--67}\BibitemShut {NoStop}%
\bibitem [{\citenamefont {Anschuetz}\ \emph {et~al.}(2019)\citenamefont {Anschuetz}, \citenamefont {Olson}, \citenamefont {Aspuru-Guzik},\ and\ \citenamefont {Cao}}]{anschuetz2018variational}%
  \BibitemOpen
  \bibfield  {author} {\bibinfo {author} {\bibfnamefont {E.}~\bibnamefont {Anschuetz}}, \bibinfo {author} {\bibfnamefont {J.}~\bibnamefont {Olson}}, \bibinfo {author} {\bibfnamefont {A.}~\bibnamefont {Aspuru-Guzik}}, \ and\ \bibinfo {author} {\bibfnamefont {Y.}~\bibnamefont {Cao}},\ }\bibfield  {title} {\enquote {\bibinfo {title} {Variational quantum factoring},}\ }in\ \href@noop {} {\emph {\bibinfo {booktitle} {Quantum Technology and Optimization Problems}}},\ \bibinfo {editor} {edited by\ \bibinfo {editor} {\bibfnamefont {S.}~\bibnamefont {Feld}}\ and\ \bibinfo {editor} {\bibfnamefont {C.}~\bibnamefont {Linnhoff-Popien}}}\ (\bibinfo  {publisher} {Springer International Publishing},\ \bibinfo {address} {Cham},\ \bibinfo {year} {2019})\ pp.\ \bibinfo {pages} {74--85}\BibitemShut {NoStop}%
\bibitem [{\citenamefont {Mohammad}, \citenamefont {Pivoluska},\ and\ \citenamefont {Plesch}(2024)}]{plesch2024}%
  \BibitemOpen
  \bibfield  {author} {\bibinfo {author} {\bibfnamefont {I.~A.}\ \bibnamefont {Mohammad}}, \bibinfo {author} {\bibfnamefont {M.}~\bibnamefont {Pivoluska}}, \ and\ \bibinfo {author} {\bibfnamefont {M.}~\bibnamefont {Plesch}},\ }\bibfield  {title} {\enquote {\bibinfo {title} {Meta-optimization of resources on quantum computers},}\ }\href {\doibase 10.1038/s41598-024-59618-y} {\bibfield  {journal} {\bibinfo  {journal} {Sci. Rep.}\ }\textbf {\bibinfo {volume} {14}},\ \bibinfo {pages} {10312} (\bibinfo {year} {2024})}\BibitemShut {NoStop}%
\bibitem [{\citenamefont {If}\ and\ \citenamefont {only If~(Iff)~Technologies}(2021)}]{gsgmorph}%
  \BibitemOpen
  \bibfield  {author} {\bibinfo {author} {\bibnamefont {If}}\ and\ \bibinfo {author} {\bibnamefont {only If~(Iff)~Technologies}},\ }\href@noop {} {\enquote {\bibinfo {title} {Gsgmorph},}\ }\bibinfo {howpublished} {\url{https://github.com/IffTech/GSG-Morph}} (\bibinfo {year} {2021})\BibitemShut {NoStop}%
\bibitem [{\citenamefont {Szegedy}(2019)}]{szegedy2019qaoa}%
  \BibitemOpen
  \bibfield  {author} {\bibinfo {author} {\bibfnamefont {M.}~\bibnamefont {Szegedy}},\ }\bibfield  {title} {\enquote {\bibinfo {title} {What do qaoa energies reveal about graphs?}}\ }\href {https://arxiv.org/abs/1912.12277} {\bibfield  {journal} {\bibinfo  {journal} {arXiv:1912.12277}\ } (\bibinfo {year} {2019})}\BibitemShut {NoStop}%
\bibitem [{\citenamefont {Bonati}(2014)}]{2014P}%
  \BibitemOpen
  \bibfield  {author} {\bibinfo {author} {\bibfnamefont {C.}~\bibnamefont {Bonati}},\ }\bibfield  {title} {\enquote {\bibinfo {title} {The peierls argument for higher dimensional ising models},}\ }\href {\doibase 10.1088/0143-0807/35/3/035002} {\bibfield  {journal} {\bibinfo  {journal} {Eur. J. Phys.}\ }\textbf {\bibinfo {volume} {35}},\ \bibinfo {pages} {035002} (\bibinfo {year} {2014})}\BibitemShut {NoStop}%
% \bibitem [{\citenamefont {Istrail}(2000)}]{10.1145/335305.335316}%
%   \BibitemOpen
%   \bibfield  {author} {\bibinfo {author} {\bibfnamefont {S.}~\bibnamefont {Istrail}},\ }\bibfield  {title} {\enquote {\bibinfo {title} {Statistical mechanics, three-dimensionality and np-completeness: I. universality of intracatability for the partition function of the ising model across non-planar surfaces},}\ }in\ \href {\doibase 10.1145/335305.335316} {\emph {\bibinfo {booktitle} {Proc. 32nd Annu. ACM Symp. Theory Comput.}}},\ \bibinfo {series and number} {STOC '00}\ (\bibinfo  {publisher} {Association for Computing Machinery},\ \bibinfo {address} {New York, NY, USA},\ \bibinfo {year} {2000})\ p.\ \bibinfo {pages} {87–96}\BibitemShut {NoStop}%
% \bibitem [{\citenamefont {Barahona}(1982)}]{Barahona_1982}%
%   \BibitemOpen
%   \bibfield  {author} {\bibinfo {author} {\bibfnamefont {F.}~\bibnamefont {Barahona}},\ }\bibfield  {title} {\enquote {\bibinfo {title} {On the computational complexity of ising spin glass models},}\ }\href {\doibase 10.1088/0305-4470/15/10/028} {\bibfield  {journal} {\bibinfo  {journal} {J. Phys. A: Math. Gen.}\ }\textbf {\bibinfo {volume} {15}},\ \bibinfo {pages} {3241--3253} (\bibinfo {year} {1982})}\BibitemShut {NoStop}%
\end{thebibliography}


\end{document}